\documentclass[pre,twocolumn,nofootinbib]{revtex4-1}
\usepackage{amsmath,amssymb}
\usepackage[dvipdfmx]{graphicx}
\usepackage[dvipdfmx]{hyperref}
\usepackage{bm}
\usepackage{mathrsfs}
\usepackage{braket}
\usepackage{comment}
\usepackage{color}
\newcommand{\mJ}{\mathcal{J}}
\newcommand{\mH}{\mathcal{H}}

\definecolor{forest}{rgb}{ .15, .55, .15}

\makeatletter

\begin{document}

\title{Eigenvalue analysis of stress-strain curve of two-dimensional amorphous solids of dispersed frictional grains with finite shear strain}
\author{Daisuke Ishima$^{1}$, Kuniyasu Saitoh$^{2}$, Michio Otsuki$^{3}$, and Hisao Hayakawa$^{1}$}
\affiliation{
$^{1}$Yukawa Institute for Theoretical Physics, Kyoto University, \\
Kitashirakawa-oiwake cho, Sakyo-ku, Kyoto 606-8502, Japan,\\
$^{2}$Department of Physics, Faculty of Science, Kyoto Sangyo University,\\
Motoyama, Kamigamo, Kita-ku, Kyoto 603-8555, Japan,\\
$^{3}$Graduate School of Engineering Science, Osaka University, Toyonaka, Osaka 560-8531, Japan
}
\date{\today}
\begin{abstract}
The stress-strain curve of two-dimensional frictional dispersed grains interacting with a harmonic potential without considering the dynamical slip under a finite strain is determined by using eigenvalue analysis of the Hessian matrix. 
After the configuration of grains is obtained, 
the stress-strain curve based on the eigenvalue analysis is in almost perfect agreement with that obtained by the simulation, 
even if there are plastic deformations caused by stress avalanches. 
Unlike the naive expectation, the eigenvalues in our model do not indicate any precursors to the stress-drop events.
\end{abstract}
\maketitle

\section{Introduction\label{intro}}

Amorphous materials consisting of dispersed repulsive grains, such as powders, colloids, bubbles, and emulsions, behave as fragile solids above jamming density~\cite{Jaeger96,Durian95,Wyart05,Baule18,Liu98}.
When we consider a response of such materials to an applied strain $\gamma$, the rigidity $G$ is independent of the strain in the linear response regime, whereas it exhibits softening in the nonlinear regime~\cite{Nagamanasa14,Knowlton14,Kawasaki16,Leishangthem17,Bohy17,Ozawa18,Clark18,Boschan19,Singh20,Otsuki21}.
Above the yielding points, there are some plastic events, such as stress avalanches in the collection of grains.

The Hessian matrix determined by the configuration of grains is commonly used for amorphous solids consisting of frictionless grains ~\cite{Wyart05,WyarLettt05,Ellenbroek06, Lerner16, Gartner16, Bonfanti19, Baule18, Morse20}.
To determine the rigidity,  eigenvalue analysis of the Hessian matrix~\cite{DeGiuli14, Mizuno16, Maloney04, Maloney06, Lemaitre06, Zaccone11, Palyulin18}, is commonly used, but we have only obtained semi-quantitative agreements between the theory and simulation so far~\cite{ Zaccone11, Palyulin18}. 
Recently, some researchers examined the applicability of the instantaneous normal mode analysis to systems that are not fully equilibrated and found qualitative agreement between the analysis and simulation~\cite{Oyama21,Kriuchevskyi22}, 
although they could not get quantitatively accurate results. 
Some studies have suggested that the decrement of the non-zero smallest eigenvalue of the Hessian matrix with the strain is a precursor of an avalanche or stress drop near a critical strain $\gamma_{c}$ ~\cite{Maloney04, Maloney06, Manning11, Dasgupta12, Ebrahem20}.
Correspondingly, some studies indicated that the rigidity $G$ and the stress $\sigma$ should behave as $G-G_{\rm reg}\propto - 1/\sqrt{\gamma_c-\gamma}$ and $\sigma-\sigma_{\rm reg} \propto \sqrt{\gamma_c-\gamma}$ near $\gamma_c$, where $G_{\rm reg}$ and $\sigma_{\rm reg}$ are the regular parts of the rigidity and stress conversing to constants at $\gamma_c$, respectively~\cite{Maloney04, Maloney06, Karmakar10PRL,Richard20}.

In general, the frictional force between the grains cannot be ignored in physical situations. 
Because the frictional force generally depends on the contact history, Hessian analysis is not applicable to such systems.
Thus, Chattoraj et al. adopted the Jacobian matrix instead of the Hessian matrix to discuss the stability of the configuration of frictional grains under strain~\cite{Chattoraj19}.
They performed eigenvalue analysis under athermal quasi-static shear processes and determined the existence of oscillatory instability originating from interparticle friction at a certain strain \cite{Chattoraj19, Chattoraj19E, Charan20}.
Moreover, some studies have performed an analysis of the Hessian matrix with the aid of an effective potential for frictional grains~\cite{Somfai07, Henkes10}.
Recently, Liu et al. suggested that Hessian matrix analysis with another effective potential can be used, even if slip processes exist \cite{Liu21}.
Previous studies \cite{Somfai07,Henkes10} have reported that the friction between grains causes a continuous change in the functional form of the density of states (DOS), which differs from that of frictionless systems. 
So far, there have been few theoretical studies to determine the rigidity of the frictional grains.

In our previous study~\cite{Ishima2022}, we developed an analysis of the Jacobian matrix to determine the rigidity of two-dimensional 
amorphous solids consisting of frictional grains interacting with the Hertzian force in the linear response to an infinitesimal strain. 
In the study, we ignore the dynamic friction caused by the slip processes of contact points.
We found that there are two modes in the DOS for a sufficiently small tangential-to-normal stiffness ratio.
Rotational modes exist in the region of low-frequency or small eigenvalues, whereas translational modes exist in the region of high-frequency or large eigenvalues. 
The rigidity determined by the translational modes is in good agreement with that obtained by the molecular dynamics simulations, 
whereas the contribution of the rotational modes is almost zero. 
Nevertheless, there are several shortcomings in the previous analysis. 
(i) The analysis can be used only in the linear response regime, where the base state is not influenced by the applied strain. 
(ii) As a result, we cannot discuss the behavior of plastic deformations or avalanches of grains. 
(iii) Even if we restrict our interest to the linear response regime, we cannot include the effect of tangential contact for the preparation of the initial configuration.
(iv) We also ignored the effect of Coulomb's slip between the contacted grains~\cite{Ishima2022}.

The purpose of this study is to overcome the shortcomings of our previous study except for point (iv)~\cite{Ishima2022}.
Thus, we analyze a collection of two-dimensional grains interacting with repulsive harmonic potentials within the contact radius, 
without considering Coulomb's slip between the grains. 
Owing to the special properties of the harmonic potential, 
the eigenvalue analysis of the Jacobian matrix becomes equivalent to that of the Hessian matrix. 
Subsequently, using the eigenvalue analysis of the Hessian matrix,
we demonstrate that the theoretical rigidity under a large strain agrees with that obtained by the simulation.

The remainder of this paper is organized as follows.
In Sec.~\ref{Sec2}, we introduce the model to be analyzed in this study.
In Sec.~\ref{Sec3}, we summarize the theoretical framework for determining the rigidity of an amorphous solid consisting of frictional grains without considering Coulomb's slip process.
In Sec.~\ref{results}, we present the results of the stress-strain relation obtained using the theory formulated in Sec.~\ref{Sec3}.
We also compare the theoretical results with the simulation results to demonstrate the relevancy of our theoretical analysis.
In Sec.~\ref{concluding_remarks}, we summarize the obtained results and address future tasks to be solved.
In Appendix~\ref{App:Absence_of_2nd}, we numerically demonstrate the absence of history-dependent tangential deformations in our system.
In Appendix~\ref{app:gamma_Th}, we explain the detailed behavior of the eigenvalue near the stress-drop points.
In Appendix~\ref{AppHessian}, we present some detailed properties of the Hessian matrix in a harmonic system.
In Appendix~\ref{AppHarmonic}, we present some properties of the Jacobian matrix in a harmonic system and demonstrate its equivalency to the Hessian matrix.
Appendix~\ref{app:rigidity} presents the detailed properties of rigidity.

\section{Model \label{Sec2}}

Our system contains $N$ frictional circular disks embedded in a two-dimensional space.
To prevent the system from crystallization, it contains an equal number of grains with diameters $d$ and $d/1.4$ \cite{Luding01}.
We assume that the mass of grain $i$ is proportional to $d_i^2$, where $d_i$ is the diameter of $i$-th grain.
For later convenience, we introduce $m$ as the mass of grain having a diameter $d$.
In this study, $x_{i}, y_{i}$, and $\theta_{i}$ denote $x$ and $y$ components of the position of $i$-th grain, and the rotational angle of the $i$-th grain, respectively.
We introduce the generalized coordinates of the $i$-th grain as follow: 
\begin{equation}\label{def_q}
\bm{q}_{i}:=(\bm{r}_{i}^{\textrm{T}},\ell_{i})^{\textrm{T}}, 
\end{equation}
where $\bm{r}_{i}:=(x_{i},y_{i})^{\textrm{T}}$, $\ell_{i}:=d_{i}\theta_{i}/2$, and the superscript $\textrm{T}$ denotes the transposition.

Let the force, and $z$-component of the torque acting on the $i$-th grain be $\bm{F}_{i}:=(F_{i}^{x},F_{i}^{y})^{\textrm{T}}$ and $T_{i}$, respectively.
Then, the equations of motion of $i$-th grain are expressed as
\begin{align}
m_{i}\frac{d^{2}\bm{r}_{i}}{dt^{2}} &=  \bm{F}_{i}, \label{tra}\\
I_{i}\frac{d^{2}\theta_{i}}{dt^{2}} &= T_{i}, \label{rot}
\end{align}
with mass $m_i$ and momentum of inertia $I_i:=m_i d_i^2/8$ of $i$-th grain.
In a system without volume forces, such as gravity, we can write
\begin{align}
\bm{F}_{i}&=\sum_{j\neq i} \bm{f}_{ij} - m_{i}\eta_{D}\dot{\bm{r}}_{i}, \label{Fi}\\
T_{i}&=\sum_{j\neq i}T_{ij} - I_{i}\eta_{D}\dot{\theta}_{i}, \label{Ti}
\end{align}
where we have adopted the notation $\dot{A}:=dA/dt$ for arbitrary variable $A$ such as $A=\bm{r}_i$ and ${\theta}_i$. 
Here, $\bm{f}_{ij}$ and $T_{ij}$ are the force and $z$-component of the torque acting on the $i$-th grain from the $j$-th grain, respectively.
As a simplified description of the drag terms from the background fluid, 
$\eta_D$ is a damping constant uniformly acting on grains, and
$T_{ij}$ is given by
\begin{align}
T_{ij}=-\frac{d_{i}}{2}( {n}_{ij}^{x}{f}_{ij}^{y} - {n}_{ij}^{y}{f}_{ij}^{x} ),
\label{Torque}
\end{align}
where we have introduced the normal unit vector between $i$-th and $j$-th grains as $\bm{n}_{ij}:=\bm{r}_{ij}/|\bm{r}_{ij}|:=(\bm{r}_{i}-\bm{r}_{j})/|\bm{r}_{i}-\bm{r}_{j}|$.
The force $\bm{f}_{ij}$ can be divided into the normal part $\bm{f}_{N,ij}$ and tangential part $\bm{f}_{T,ij}$ as
\begin{align}
\bm{f}_{ij}=(\bm{f}_{N,ij}+\bm{f}_{T,ij} )H(d_{ij}/2-|\bm{r}_{ij}|)
\label{fijSum}
,
\end{align}
where $d_{ij}:=d_i+d_j$ and $H(x)$ is Heaviside's step function, taking $H(x)=1$ for $x>0,$ and $H(x)=0$ otherwise.
We assume that the contact force is expressed as
\begin{align}
\bm{f}_{N,ij}:&=k_{N}\xi_{N,ij}\bm{n}_{ij} 
\label{fijn}
,\\
\bm{f}_{T,ij}:&=k_{T}\xi_{T,ij}\bm{t}_{ij} 
\label{fijt}
,
\end{align}
where $k_N$ and $k_T$ are the stiffness parameters of normal and tangential contacts, respectively.
The contact force can be derived from the harmonic potential.
In Eqs.~\eqref{fijn} and ~\eqref{fijt} we have introduced
$\xi_{N,ij}:=d_{ij}/2-|\bm{r}_{ij}|$ and
\begin{align}
\bm{\xi}_{T,ij}(t):&=\int_{C_{ij}(t')}dt'\bm{v}_{T,ij}(t')
\notag\\
&\quad
 -\left[\left( \int_{C_{ij}(t')}dt'\bm{v}_{T,ij}(t') \right)\cdot\bm{n}_{ij}(t)\right]\bm{n}_{ij}(t) ,
\label{xiT}
\end{align}
where we have used 
\begin{align}
\bm{v}_{T,ij}:=\dot{\bm{r}}_{ij}-\dot{\bm{\xi}}_{N,ij}+\bm{u}_{ij}(d_{i}\dot{\theta}_{i}+d_{j}\dot{\theta}_{j})/2
\end{align}
with $\bm{u}_{ij}:=(n_{ij}^{y}, -n_{ij}^{x})^{\textrm{T}}$, $\bm{t}_{ij}:=-\bm{\xi}_{T,ij}/|\xi_{T,ij}|$,  $\xi_{T,ij}:=|\bm{\xi}_{T,ij}|$, 
and the integration over the duration time of contact between $i$-th and $j$-th grains $\int_{C_{ij}(t')}dt'$ with the trajectory $C_{ij}(t')$ of the contact point between $i$-th and $j$-th grains at $t'$.
As shown in Appendix~\ref{App:Absence_of_2nd}, we have confirmed that the second term on the right-hand side (RHS) of Eq.~\eqref{xiT} is zero for harmonic systems, although we do not have any mathematical proof for this statement thus far.
For simplicity, we consider neither the effects of Coulomb's slip in the tangential equation of motion nor the dissipative contact force, 
where the tangential forces $\bm{f}_{T,ij}^{C}$ including the effect of Coulomb's slip process, satisfy 
\begin{align}
\bm{f}_{T,ij}^{C}
:=
\left\{
\begin{matrix}
\bm{f}_{T,ij} &(\bm{f}_{T,ij} < \mu|\bm{f}_{N,ij}|)\\
\mu|\bm{f}_{N,ij}|\bm{t}_{ij} &(\textrm{otherwise}) 
\end{matrix}
\right.
\end{align}
where $\mu$ is the friction constant.

We impose the Lees-Edwards boundary conditions \cite{Lees72,Evans08},
where the direction parallel to the shear strain is the $x$-direction.
After generating a stable grain configuration via isotropic compression starting from a random configuration by using Eqs.~\eqref{tra}-\eqref{xiT} without strain (see detail in Ref. \cite{Ishima2022}), 
we apply a step strain $\Delta\gamma$ to all grains, where $x$-coordinate of the position of the $i$-th grain is shifted by an affine displacement $\Delta x_{i}(\Delta\gamma):=\Delta\gamma y_{i}^{\textrm{FB}}(0)$.
Here, the superscript FB denotes the force-balance (FB) state at which $\bm{F}_{i}=\bm{0}$ and $T_{i}=0$ for arbitrary $i$.
As shown in Sec.~\ref{Sec3A}, the FB state is equivalent to the potential minimum for harmonic grains.
Subsequently, the system is relaxed to an FB state.
We further apply the step strain $\Delta\gamma$ associated with the subsequent relaxation process again to obtain the state at $2\Delta \gamma$.
By repeating this process, we can reach a deformed state with the strain $\gamma$.

The plastic deformations for a large $\gamma$ depend on the choice of $\Delta \gamma$~\cite{Morse20}.
Moreover, the theoretical formulation assumes $\Delta\gamma\to 0$.
Therefore, we adopt the backtracking method~\cite{Lerner09,Karmakar10PRE}.
If a plastic event is encountered under a fixed $\Delta \gamma_{\rm in}$, 
the system is restored to its original state without a plastic event.
Subsequently, we apply a new strain, $0.1\Delta \gamma_{\rm in}$, to the system.
Even if we encounter a plastic event with $0.1\Delta \gamma_{\rm in}$, we further examine the smaller step strain of $0.01\Delta \gamma_{\rm in}$.
We repeat this procedure until we reach
$\Delta\gamma<\Delta\gamma_{\rm Th}$ (see Appendix \ref{app:gamma_Th}).

We introduce the rate of nonaffine displacements for $r_i^{{\rm FB},\zeta}(\gamma)$ with $\zeta=x$ or $y$ and $\ell_i^{\rm FB}(\gamma)$ as:
\begin{align}
\frac{d\mathring r_{i}^{\zeta}(\gamma)}{d\gamma}:&=
\lim_{\Delta\gamma\to0}\frac{r_{i}^{\textrm{FB},\zeta}(\gamma+\Delta\gamma)-r_{i}^{\textrm{FB},\zeta}(\gamma)}{\Delta\gamma}-\delta_{\zeta x} y_{i}^{\textrm{FB}}(\gamma), \label{drdG}
\\
\frac{d\mathring \ell_{i}(\gamma)}{d\gamma}:&=
\lim_{\Delta\gamma\to0}\frac{\ell_{i}^{\textrm{FB}}(\gamma+\Delta\gamma)-\ell_{i}^{\textrm{FB}}(\gamma)}{\Delta\gamma}
. \label{dldG}
\end{align}

Our system is characterized by the generalized coordinate $\bm{q}(\gamma):=(\bm{q}_{1}^{\textrm{T}}(\gamma), \bm{q}_{2}^{\textrm{T}}(\gamma), \cdots, \bm{q}_{N}^{\textrm{T}}(\gamma))^{\textrm{T}}$.
The configuration in the FB state at strain $\gamma$ is denoted by $\bm{q}^{\textrm{FB}}(\gamma):=((\bm{q}_{1}^{\textrm{FB}}(\gamma))^{\textrm{T}}, (\bm{q}_{2}^{\textrm{FB}}(\gamma))^{\textrm{T}},\cdots,(\bm{q}_{N}^{\textrm{FB}}(\gamma))^{\textrm{T}})^{\textrm{T}}$.
The shear stress $\sigma(\gamma)$ at $\bm{q}^{\textrm{FB}}(\gamma)$ for one sample is given by:
\begin{align}
\sigma(\bm{q}^{\textrm{FB}}(\gamma))=
-\frac{1}{2L^{2}} \sum_{i}\sum_{j>i} 
&\left[ f_{ij}^{x}(\bm{q}^{\textrm{FB}}(\gamma)) r_{ij}^{y}(\bm{q}^{\textrm{FB}}(\gamma))
\right.
\nonumber \\
&\ +\left. f_{ij}^{y}(\bm{q}^{\textrm{FB}}(\gamma)) r_{ij}^{x}(\bm{q}^{\textrm{FB}}(\gamma))
\right]
\label{Sigma}
.
\end{align}
The rigidity $g$ for one sample is defined as
\begin{align}
g&:= \left. \frac{d\sigma(\bm{q}(\gamma))}{d\gamma} \right|_{\bm{q}(\gamma)=\bm{q}^{\textrm{FB}}(\gamma)}
\label{g}
,
\end{align}
where the differentiation on the RHS of Eq. \eqref{g} is defined as follows:
\begin{align}
&\left. \frac{d\sigma(\bm{q}(\gamma))}{d\gamma} \right|_{\bm{q}(\gamma)=\bm{q}^{\textrm{FB}}(\gamma)}
\nonumber \\
&\quad\quad\quad:=\lim_{\Delta \gamma\to 0}
\frac{\sigma(\bm{q}^{\textrm{FB}}(\gamma+\Delta\gamma)) - \sigma(\bm{q}^{\textrm{FB}}(\gamma))}{\Delta\gamma}
\label{Gnm}
.
\end{align}
In the numerical calculation, we use a non-zero but sufficiently small $\Delta \gamma$ for the evaluation of $g$.
Then, the averaged rigidity $G$ is defined as
\begin{align}
G&:=\Braket{ g }
\label{Gdef}
,
\end{align}
where $\langle \cdot \rangle$ is the ensemble average.

\section{Theoretical Analysis \label{Sec3}}
In this section, we introduce Hessian matrix in Sec.~\ref{Sec3A} and theoretical expressions of rigidity in Sec.~\ref{Sec3B}.

\subsection{Hessian matrix for frictional grains}\label{Sec3A}

Because the Hessian matrix is equivalent to the Jacobian matrix for harmonic grains, as shown in  Appendices~\ref{Hessian=Jacobian} and \ref{AppHkn2}, 
in this study, we adopt the Hessian matrix ($\mH$), where the element is given by~\cite{Saitoh19}:
\begin{align}
\mH_{ij}^{\alpha\beta}:=\left. \frac{\partial^{2} \delta e_{ij} (\bm{q}(\gamma))}{\partial q_{i}^{\alpha}\partial q_{j}^{\beta}} \right|_{\bm{q}(\gamma)=\bm{q}^{\textrm{FB}}(\gamma)},
\label{Jacobian}
\end{align}
where $\alpha$ and $\beta$ are any of $x,y$ and $\ell$, while $i$ and $j$ express the grain indices.
Here, we have introduced the effective potential energy $\delta e_{ij}$ between the contacted $i$-th and $j$-th grains as:
\begin{align}
\delta e_{ij}:=\frac{k_{N}}{2}(\delta\bm{r}_{ij}\cdot\bm{n}_{ij})^{2} + \frac{k_{T}}{2}\delta\bm{r}_{ij,\perp}^{2} ,
\label{potential}
\end{align}
where $\delta\bm{r}_{ij,\perp}$ is defined as
\begin{align}
\delta\bm{r}_{ij,\perp}:=\delta\bm{r}_{ij}-(\delta\bm{r}_{ij}\cdot\bm{n}_{ij})\bm{n}_{ij}-\delta\bm{\ell}_{ij}\times\bm{n}_{ij} 
\label{tangential}
\end{align}
with
\begin{align}
\delta\bm{r}_{ij}:&=\delta\bm{r}_{i}-\delta\bm{r}_{j},\\
\delta\bm{\ell}_{ij}:&=(\delta\ell_{i}+\delta\ell_{j})\bm{e}_{z} 
\end{align}
under the virtual displacements $\delta\bm{r}_i$ and $\delta\ell_i$ from the FB state at $\bm{r}_i^{\rm FB}$ and $\ell_i^{\rm FB}$, respectively.

The Hessian matrix introduced in Eq.~\eqref{Jacobian} can be written as
\begin{align}
\mH=
\left[
\begin{matrix}
\mH_{11} & \cdots & \mH_{1i} & \cdots & \mH_{1j} & \cdots & \mH_{1N}\\
\vdots & \ddots & \vdots &   & \vdots &  &\vdots \\
\mH_{i1} & \cdots & \mH_{ii} & \cdots & \mH_{ij} & \cdots & \mH_{iN}\\
\vdots &  & \vdots & \ddots & \vdots &  &\vdots \\
\mH_{j1} & \cdots & \mH_{ji} & \cdots & \mH_{jj} & \cdots & \mH_{jN}\\
\vdots &  & \vdots &   & \vdots & \ddots &\vdots \\
\mH_{N1} & \cdots & \mH_{Ni} & \cdots & \mH_{Nj} & \cdots & \mH_{NN}
\end{matrix}
\right]
,
\label{defJ}
\end{align}
where $\mH_{ij}$ is a $3\times3$ submatrix of the Hessian $\mH$ for a pair of grains $i$ and $j$ satisfying:
\begin{align}
\mH_{ij}=
\left[
\begin{matrix}
\mH_{ij}^{xx} & \mH_{ij}^{xy} & \mH_{ij}^{x\ell} \\
\mH_{ij}^{yx} & \mH_{ij}^{yy} & \mH_{ij}^{y\ell} \\
\mH_{ij}^{\ell x} & \mH_{ij}^{\ell y} & \mH_{ij}^{\ell\ell}
\end{matrix}
\right]
\label{33J}
.
\end{align}
See  Appendix \ref{Hessian=Jacobian} for an explicit expression of each component of the Hessian matrix.
Note that $\mH_{ij}^{\alpha\beta}=0$ if the $i$-th and $j$-th grains are not in contact with each other.

Because $\mH$ is a real symmetric matrix, its eigenvalues and eigenvectors are also real. 
Using the decomposition of the potential, the Hessian matrix can be divided into
\begin{align}
\mH_{ij}^{\alpha\beta}=\mH_{N,ij}^{\alpha\beta}+\mH_{T,ij}^{\alpha\beta}
\label{sepNT}
,
\end{align}
where
\begin{align}
\mH_{N,ij}^{\alpha\beta}:&=\left. \frac{\partial^{2} \delta e_{N,ij}^{\alpha}(\bm{q}(\gamma))}{\partial q_{i}^{\alpha}\partial q_{j}^{\beta}} \right|_{\bm{q}(\gamma)=\bm{q}^{\textrm{FB}}(\gamma)},\\
\mH_{T,ij}^{\alpha\beta}:&=\left. \frac{\partial^{2} \delta e_{T,ij}^{\alpha}(\bm{q}(\gamma))}{\partial q_{i}^{\alpha}\partial q_{j}^{\beta}} \right|_{\bm{q}(\gamma)=\bm{q}^{\textrm{FB}}(\gamma)}
\end{align}
for $i\neq j$ and
\begin{align}
\mH_{N,ij}^{\alpha\beta}:&=
\left. \frac{\partial^{2} \delta e_{N,ik}^{\alpha}(\bm{q}(\gamma))}{\partial q_{i}^{\alpha}\partial q_{i}^{\beta}} \right|_{\bm{q}(\gamma)=\bm{q}^{\textrm{FB}}(\gamma)},\\
\mH_{T,ij}^{\alpha\beta}:&=
\left. \frac{\partial^{2} \delta e_{T,ik}^{\alpha}(\bm{q}(\gamma))}{\partial q_{i}^{\alpha}\partial q_{i}^{\beta}} \right|_{\bm{q}(\gamma)=\bm{q}^{\textrm{FB}}(\gamma)}
\end{align}
for $i=j$.
Here, we have introduced 
\begin{align}
\delta e_{N,ij}:&=\frac{k_{N}}{2}(\delta\bm{r}_{ij}\cdot\bm{n}_{ij})^{2},    \\
\delta e_{T,ij}:&=\frac{k_{T}}{2}\delta\bm{r}_{ij,\perp}^{2} .
\end{align}
To determine the explicit expression of each component of the Hessian matrix, refer to Appendix~\ref{Hessian=Jacobian}.

The eigenequation of the Hessian matrix $\mH$ is given by
\begin{align}
\mH\ket{\Phi_{n}}&= \lambda_{n}\ket{\Phi_{n}}, \label{REieq}
\end{align}
where $\ket{\Phi_{n}}$ is the right eigenvector corresponding to the $n$-th eigenvalue $\lambda_{n}$ of $\mH$.
Because the Hessian matrix is a real symmetric matrix, its left eigenequation is equivalent to its right eigenequation. 
Such properties remain unchanged even under the Lees-Edwards boundary conditions (see Appendix \ref{app:eff_boundary}).
If all eigenstates are non-degenerate, $\ket{\Phi_n}$ satisfies the orthonormal relation $\braket{\Phi_m|\Phi_n}=\delta_{mn}$ with normalization $\braket{\Phi_n|\Phi_n}=1$, 
where $\braket{\Phi_{n}|\Phi_{n}}:=\sum_{i=1}^{N}\sum_{\alpha=x,y,\ell}(\Phi_{n,i}^{\alpha})^{2}$.

\subsection{Expressions of the rigidity via eigenmodes \label{Sec3B}}

In this subsection, we consider the rigidity $g$ introduced in Eq.~\eqref{Gdef}.
See Appendix~\ref{app:rigidity} for the detailed properties of $g$. 

Let us introduce $\tilde{\bm{F}}_i := (\tilde{F}^x_i , \tilde{F}^y_i , \tilde{F}^\ell_i)^{\rm T} := (F^x_i, F^y_i,2T_i/d_i)^{\rm T}$ and $\ket{\tilde F(\bm{q}(\gamma))}$ as
\begin{align}
\ket{\tilde F(\bm{q}(\gamma))}:=[ \tilde{\bm{F}}_{1}^{\textrm{T}}(\bm{q}(\gamma)), \tilde{\bm{F}}_{2}^{\textrm{T}}(\bm{q}(\gamma)), \cdots, \tilde{\bm{F}}_{N}^{\textrm{T}}(\bm{q}(\gamma)) ]^{\textrm{T}}.
\end{align}
Because the FB state is the minimum state of the potential energy, as shown in Appendix~\ref{Hessian=Jacobian},
$\ket{\tilde{F}(\bm{q}(\gamma))}|_{\bm{q}(\gamma)=\bm{q}^{\textrm{FB}}(\gamma)}$ satisfies
\begin{align}
\left. \ket{\tilde F(\bm{q}(\gamma))} \right|_{ \bm{q}(\gamma)=\bm{q}^{\textrm{FB}}(\gamma) }=\left. \frac{d\ket{\tilde F(\bm{q}(\gamma))}}{d\gamma} \right|_{\bm{q}(\gamma)=\bm{q}^{\textrm{FB}}(\gamma)}=\ket{0},
\end{align}
where $\ket{0}$ is the ket vector containing $0$ for all components.

Introducing
\begin{align}
\Ket{\frac{d\mathring q}{d\gamma}}
:=
\left[
\frac{d\mathring r_{1}^{x}}{d\gamma},
\frac{d\mathring r_{1}^{y}}{d\gamma},
\frac{d\mathring \ell_{1}}{d\gamma},
\cdots,
\frac{d\mathring r_{N}^{x}}{d\gamma},
\frac{d\mathring r_{N}^{y}}{d\gamma},
\frac{d\mathring \ell_{N}}{d\gamma}
\right]^{\textrm{T}}
,
\label{NonAff}
\end{align}
one can write 
 \begin{align}
\left. \frac{d\ket{\tilde F(\bm{q}(\gamma))}}{d\gamma} \right|_{\bm{q}(\gamma)=\bm{q}^{\textrm{FB}}(\gamma)} =-\ket{\Xi} +\tilde\mH\Ket{\frac{d\mathring q}{d\gamma}}
\label{dFdG2}
,
\end{align}
where we have used Eqs. \eqref{drdG} and \eqref{dldG}. 
The first and second terms on the RHS of Eq. \eqref{dFdG2} represent
the strain derivatives of the forces for the contributions from the affine and nonaffine displacements, respectively.
In Eq.~\eqref{dFdG2} we have introduced $\ket{\Xi}$, which is defined as:
\begin{align}
\ket{\Xi}:=\sum_{j}
\left[
\begin{matrix}
\mH_{N,j1}^{xx}r_{1j}^{y}\\
\mH_{N,j1}^{xy}r_{1j}^{y}\\
\mH_{N,j1}^{x\ell}r_{1j}^{y}\\
\vdots\\
\mH_{N,jN}^{xx}r_{Nj}^{y}\\
\mH_{N,jN}^{xy}r_{Nj}^{y}\\
\mH_{N,jN}^{x\ell}r_{Nj}^{y}
\end{matrix}
\right]
.
\end{align}
We have used $\tilde\mH$ in Eq. \eqref{dFdG2}, which is defined as
\begin{align}
\tilde\mH_{ii}^{\alpha\beta}
&:=
\left\{
\begin{matrix}
-\mH_{ii}^{\ell x} & (\alpha=\ell,\ \beta=x) \\
-\mH_{ii}^{\ell y} & (\alpha=\ell,\ \beta=y) \\
\mH_{ii}^{\alpha\beta} & (\textrm{otherwise})
\end{matrix}
\right.
\label{tildeJ1}
\end{align}
and
\begin{align}
\tilde\mH_{ij}^{\alpha\beta}
&:=
\left\{
\begin{matrix}
-\mH_{ij}^{x\ell} & (\alpha=x,\ \beta=\ell) \\
-\mH_{ij}^{y\ell} & (\alpha=y,\ \beta=\ell) \\
\mH_{ij}^{\alpha\beta} & (\textrm{otherwise})
\end{matrix}
\right.
\label{tildeJ2}
\end{align}
for $i\neq j$.
Note that $\mH_{T,ij}^{\alpha\beta}$ or $\tilde\mH_{T,ij}^{\alpha\beta}$ does not affect $\ket{\Xi}$, 
because the affine displacements are instantaneously applied to the system as a step strain.
Thus, the integral interval of the tangential displacement during the affine deformation is zero.

Expanding the nonaffine displacements by the eigenvectors of $\tilde\mH$ and using the fact that the left-hand side of Eq. \eqref{dFdG2} is zero, we obtain
\begin{align}
\Ket{\frac{d\mathring q}{d\gamma}}=\sideset{}{^{'}}\sum_{n} \frac{\braket{\tilde{\Phi}_{n}|\Xi}}{\tilde{\lambda}_{n}}\ket{\tilde{\Phi}_{n}},
\label{eqEiExp}
\end{align}
where $\tilde{\lambda}_{n}$ and $\ket{\tilde{\Phi}_{n}}$ are the $n$-th eigenvalues of $\tilde\mH$, and the eigenvector corresponding to $\tilde{\lambda}_n$, respectively.
Here, $\sum_{n}\nolimits'$ on the RHS of Eq. \eqref{eqEiExp} excludes low-frequency modes for $\tilde\lambda_{n}t_{0}^{2}/m\leq10^{-12}$ to maintain the numerical accuracy.
Note that $\ket{\tilde{\Phi}_n}$ satisfies the orthonormal relation $\braket{\tilde{\Phi}_m|\tilde{\Phi}_n}=\delta_{mn}$, if all eigenstates are non-degenerate.
The expression for $d\mathring q/d\gamma$ in Eq.~\eqref{eqEiExp} leads to a discontinuous change in $d\mathring q/d\gamma$ at a critical strain $\gamma_c$ for a plastic event 
because the eigenvectors and eigenvalues are discontinuously changed at this point.

The rigidity is decomposed into two parts:
\begin{align}
g:=g_{\textrm{A}}+g_{\textrm{NA}} ,
\label{expG}
\end{align}
where $g_{\textrm{A}}$ and $g_{\textrm{NA}}$ are the rigidities corresponding to the affine and nonaffine displacements, respectively, for one sample.
With the aid of Eqs. \eqref{Sigma}, \eqref{Gdef}, and \eqref{tildeJ2}, the expressions for $g_{\textrm{A}}$ and $g_{\textrm{NA}}$ can be obtained as:
\begin{align}
g_{\textrm{A}}
:&=
\frac{1}{4 L^{2}}
\sum_{i,j(i\neq j)}
r_{ij}^{y}
\left[
r_{ij}^{y}\mH_{N,ji}^{xx}
+
r_{ij}^{x}\mH_{N,ji}^{yx}
\right]
\label{GAff}
, \\
g_{\textrm{NA}}
:&=
\frac{1}{4 L^{2}}
\sum_{i,j(i\neq j)}
\Biggl[
\sum_{\zeta=x,y}
 \left( r_{ij}^{y} \tilde\mH_{ij}^{x\zeta}  + r_{ij}^{x} \tilde\mH_{ij}^{y\zeta}  \right) \frac{d\mathring r_{ij}^{\zeta}}{d\gamma} \nonumber \\
&\quad\quad\quad\quad\quad\quad\quad\quad\quad
-\left( r_{ij}^{y} \tilde\mH_{ij}^{x\ell}  + r_{ij}^{x} \tilde\mH_{ij}^{y\ell} \right) \frac{d\mathring \ell_{ij}}{d\gamma}
\Biggl]
,
\label{GNAff}
\end{align}
where we have introduced
\begin{align}
\frac{d\mathring r_{ij}^{\zeta}}{d\gamma}:&=\frac{d\mathring r_{i}^{\zeta}}{d\gamma} - \frac{d\mathring r_{j}^{\zeta}}{d\gamma}, \\
\frac{d\mathring \ell_{ij}}{d\gamma}:&=\frac{d\mathring \ell_{i}}{d\gamma} + \frac{d\mathring \ell_{j}}{d\gamma}.
\end{align}

Substituting Eq. \eqref{eqEiExp} into Eq. \eqref{GNAff}, $g_{\textrm{NA}}$ can be rewritten as
\begin{align}
g_{\textrm{NA}}
&=
-
\frac{1}{L^{2}}
\sideset{}{^{'}}\sum_{n}\frac{\braket{\tilde{\Phi}_{n}|\Xi}\braket{\Theta|\tilde{\Phi}_{n}}}{\tilde\lambda_{n}}
\label{GNAffv2}
,
\end{align}
where we have introduced
\begin{widetext}
\begin{align}
\bra{\Theta}:=\frac{1}{2}\sum_{j}
\left[
\left( r_{1j}^{y}\tilde{\mH}_{j1}^{xx}+r_{1j}^{x}\tilde{\mH}_{j1}^{yx} \right), 
\left( r_{1j}^{y}\tilde{\mH}_{j1}^{xy}+r_{1j}^{x}\tilde{\mH}_{j1}^{yy} \right), 
\hdots, 
\left( r_{Nj}^{y}\tilde{\mH}_{jN}^{xy}+r_{Nj}^{x}\tilde{\mH}_{jN}^{yy} \right), 
\left( r_{Nj}^{y}\tilde{\mH}_{jN}^{x\ell}+r_{Nj}^{x}\tilde{\mH}_{jN}^{y\ell} \right)
\right] . 
\end{align}
The affine rigidity can be also expressed as
\begin{align}
g_{\textrm{A}}
= \frac{1}{L^2} \braket{Y|\Xi} ,
\end{align}
where
\begin{align}
\bra{Y}:=\left[ y_{1}, 0, 0, y_{2},0, 0,  \cdots, y_{N}, 0, 0 \right]
.
\end{align}

\end{widetext}

To verify the validity of the theoretical treatment, 
we introduce the theoretical stress $\sigma^{\textrm{th}}(\gamma)$ with the aid of Eq.\eqref{expG} as
\begin{align}\label{theoretical_sigma}
\sigma^{\textrm{th}}(\gamma+\Delta\gamma):=\sigma(\bm{q}^{\textrm{FB}}(\gamma))+g(\gamma)\Delta\gamma . 
\end{align}

\section{Results and discussion\label{results}}

We verify the validity of the shear modulus obtained by the eigenvalue analysis by comparing it with that obtained by the simulation.
First, we have confirmed the quantitative accuracy of our analysis to obtain the rigidity in the linear response regime of our system (see Appendix~\ref{AppHkn4}), as in Ref.~\cite{Ishima2022}.

For the numerical FB condition, we use the condition  $|\tilde F_{i}^{\alpha}|<F_{\textrm{Th}}$ for arbitrary $i$, 
where we adopt $F_{\textrm{Th}}=1.0\times10^{-14}k_Nd$ for the numerical calculation.
In our simulation, we also adopt $\eta_{D}=\sqrt{k_{N}/m}$. 
In this study, we  present the results for $k_{T}/k_{N}=1$,  the area fraction $\phi=0.90$, and $\Delta\gamma_{\rm in}=1.0\times10^{-4}$ with the ensemble averages of $30$ samples, except for Appendix \ref{AppHkn4}. 
Note that the we analyze only the systems with $\phi=0.90$ which is sufficiently larger than the jamming density for a two-dimensional system.
We ignore the effect of dissipation in the eigenequation because the velocity of each grain is sufficiently small to incur infinitesimal agitation from the FB state.
The time step used for the simulation, $\Delta t$, is set to $\Delta t=1.0\times10^{-2}t_{0}$ with $t_{0}:=\sqrt{m/k_{N}}$, and numerical integration is performed using the velocity Verlet method.

\begin{figure}[htbp]
\centering
\includegraphics[width=8cm]{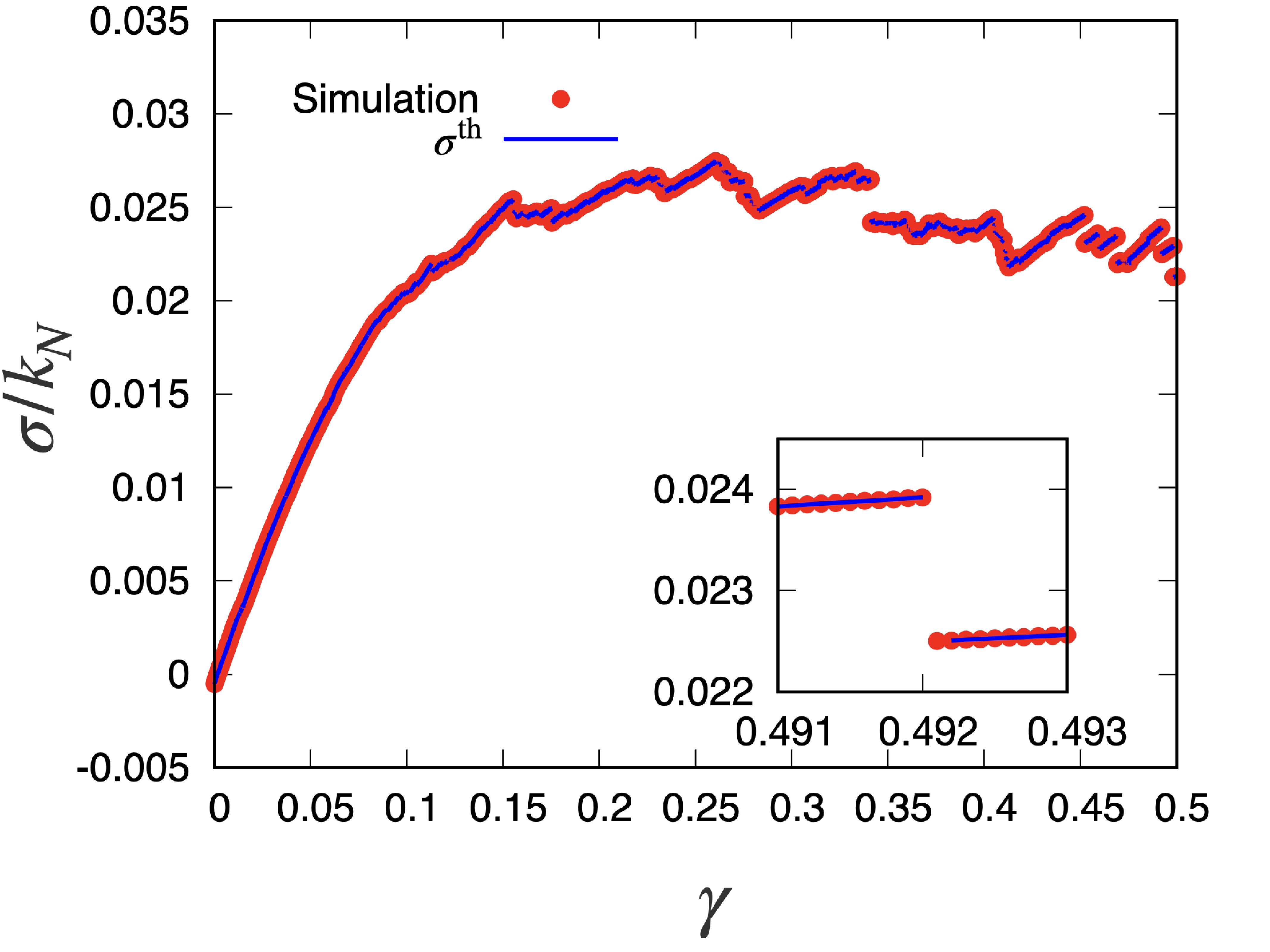}
\caption{ 
A stress-strain curve for $0\leq\gamma\leq0.5$ for one sample of the collection of grains ($N=128$), which includes the theoretical results (line) and simulation results (filled symbols) under the condition$\Delta\gamma_{\textrm{Th}}=\Delta\gamma_{\textrm{in}}=10^{-4}$. 
The inset is a close-up of the stress-strain curve in the vicinity of a stress-drop event.
}
\label{Stress}
\end{figure}

\begin{figure}[htbp]
\centering
\includegraphics[width=8cm]{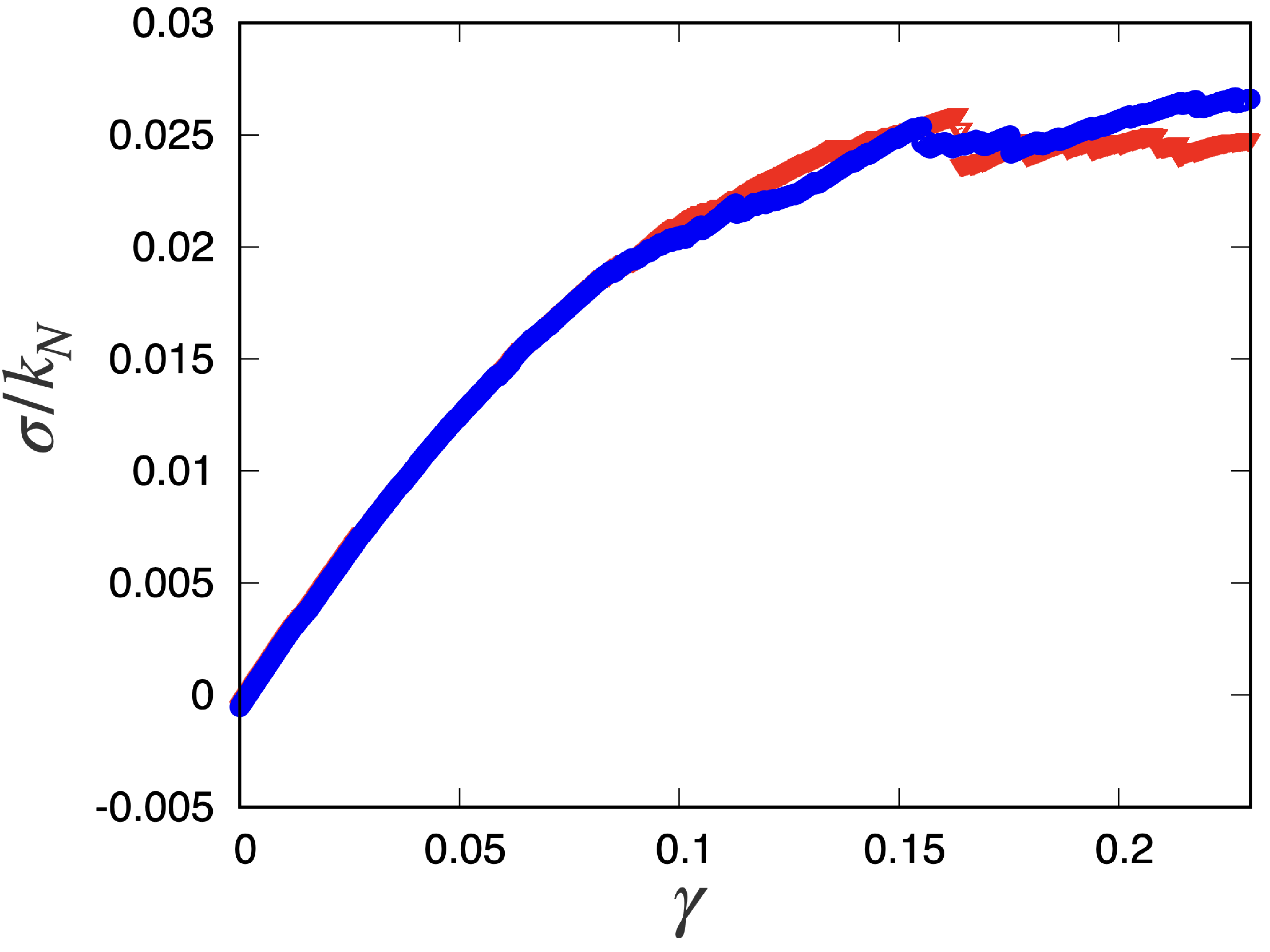}
\caption{
Stress-strain relations for
$\Delta\gamma_{\textrm{Th}}=1.0\times10^{-4}$ (blue circles) and $\Delta\gamma_{\textrm{Th}}=1.0\times10^{-8}$ (red triangles) with $N=128$.}
\label{StressCmpGammaTh}
\end{figure}

Now, let us consider a nonlinear regime in which there are many plastic events caused by stress avalanches.
Here, we regard an event as plastic if the condition (i) $\sigma(\gamma)-\sigma(\gamma-\Delta\gamma)<0$
or (ii) $G(\gamma-\Delta\gamma)-G(\gamma)>1.0\times10^{-2}k_{N}$ is satisfied.

Figure~\ref{Stress} shows a typical example of the stress-strain curve obtained by one sample of the collection of grains based on both the simulation and eigenvalue analysis developed in the previous section under the condition $\Delta\gamma_{\rm Th}=\Delta\gamma_{\rm in}$.
It should be noted that the difference between the theoretical and simulation results is almost invisible,
even in the presence of avalanches.
However, the eigenvalue analysis cannot be used immediately after plastic events, that is, for $\gamma\approx \gamma_c$ (see the inset of Fig.~\ref{Stress}), 
because the stress is not determined by Eq.~\eqref{theoretical_sigma} immediately after a plastic event.

Figure \ref{StressCmpGammaTh} shows a comparison of the stress-strain curve 
for $\Delta\gamma_{\rm Th}=10^{-4}\Delta\gamma_{\rm in}$ (blue circles) 
with those for $\Delta\gamma_{\rm Th}=\Delta\gamma_{\rm in}$ (red triangles).
From the figure, we cannot find any $\Delta\gamma_{\rm Th}$ dependence for $\gamma<0.08$, 
but some differences for larger $\gamma$ can be observed as a result of stress avalanches.

\begin{figure}[htbp]
\centering
\includegraphics[width=7cm]{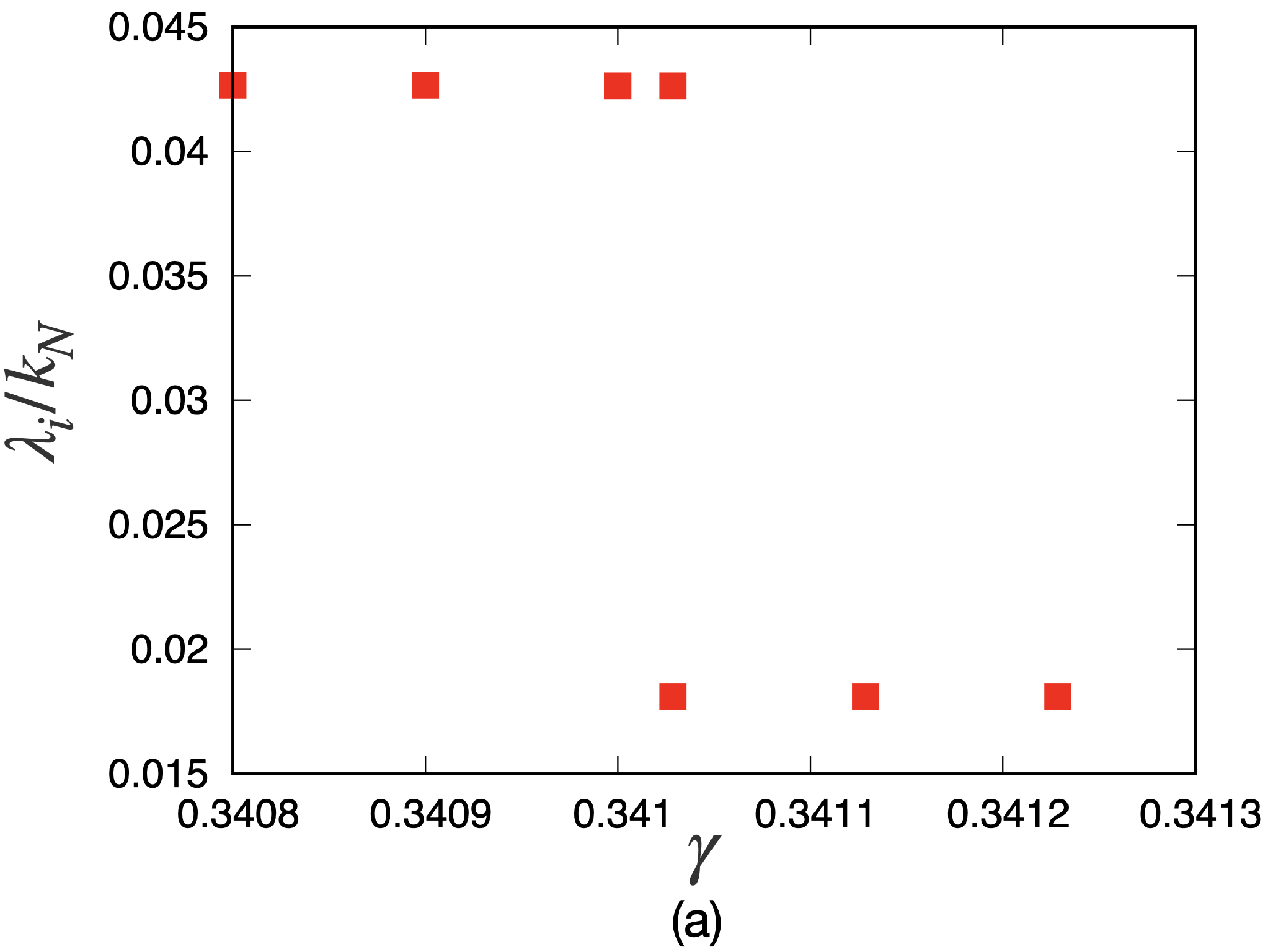}
\includegraphics[width=7cm]{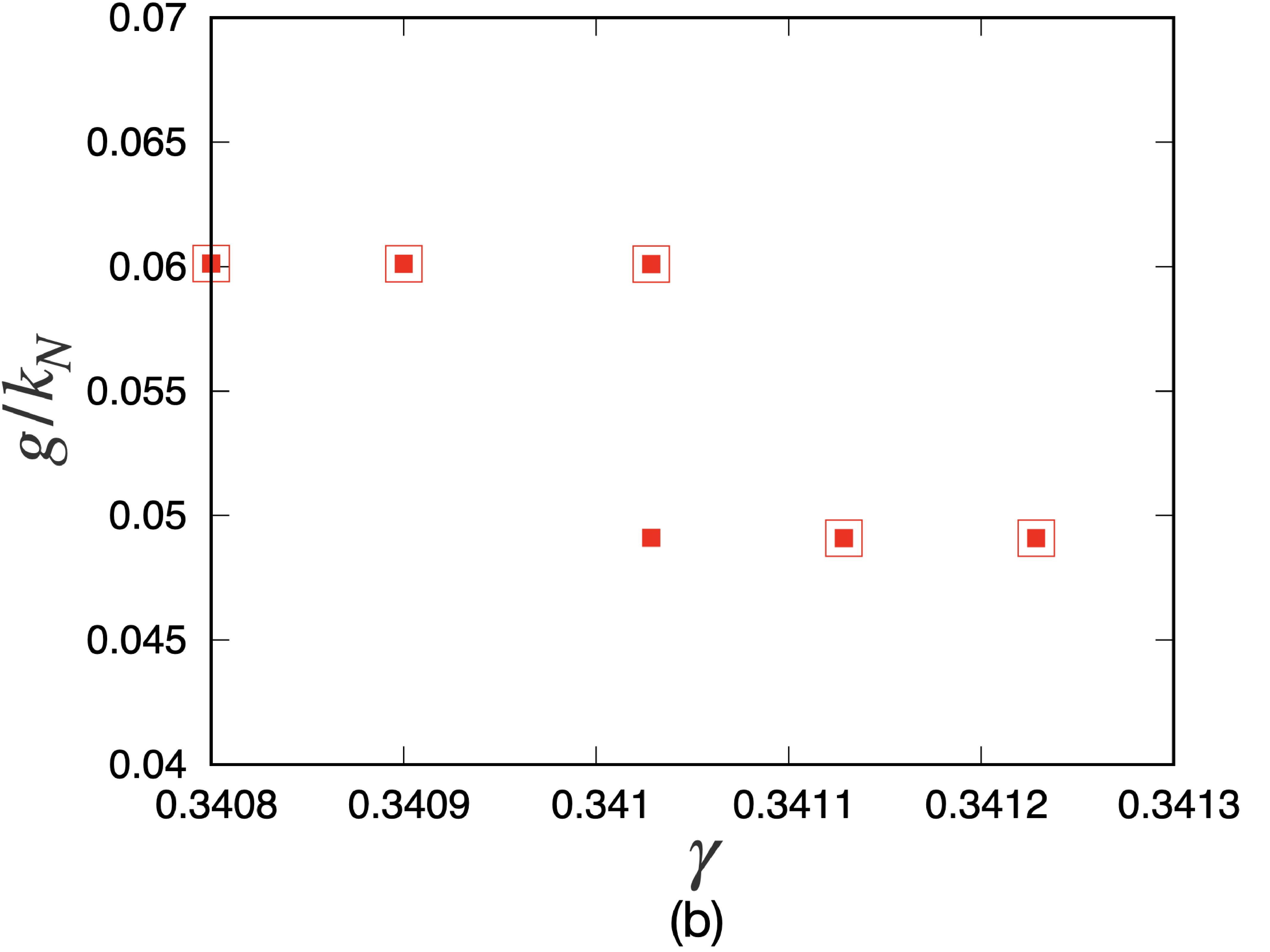}
\caption{
Plots of (a) the smallest eigenvalue except for the zero modes and  (b) the rigidity $g$ based on the eigenvalue analysis (open symbols) and numerical shear stress (filled symbols) against $\gamma$ for $0.3408\leq\gamma\leq0.3413$ and $N=128$. 
}
\label{Drop}
\end{figure}

We might expect that some precursors of a stress-drop event can be detected from the behavior of the smallest non-zero eigenvalue.
To verify this expectation, we plot the smallest non-zero eigenvalues in Fig.~\ref{Drop}(a) near a critical strain with $\Delta\gamma_{\rm Th}=10^{-8}$.
It can be observed that the eigenvalues changes discontinuously at the critical strain ($\gamma_c=0.34102862$ for $\gamma_{\rm Th}=10^{-8}$), where the critical strain converges if $\Delta\gamma_{\rm Th}<10^{-6}$ (see Appendix~\ref{app:gamma_Th} for details).
Notably, there is no precursor for the smallest eigenvalue below the critical strain, 
in contrast to Refs.~\cite{Maloney04,Maloney06,Manning11,Dasgupta12}, where non-harmonic potentials are used.
Correspondingly, we cannot find any singularity of the rigidity as $g-g_{\textrm{reg}}\sim-(\gamma_{c}-\gamma)^{-1/2}$ as in Fig.~\ref{Drop}(b) for $\gamma\lesssim \gamma_c$ predicted in Refs.~\cite{Maloney04, Maloney06, Karmakar10PRL, Richard20}. 
The absence of the precursors and singularities in our model can be understood in the form of the Hessian matrix presented in Appendix~\ref{AppHkn2}.
In non-harmonic systems, some elements of the Hessian matrix become zero as $\xi_{N,ij} \to 0$ 
when the contact between the $i$-th and $j$-th grains disappears.
This leads to the precursors and singularities \cite{Maloney04,Maloney06}.
However, in the harmonic systems, the corresponding element approaches a non-zero constant in the limit $\xi_{N,ij}\to 0$ (see Appendix~\ref{AppHkn2}), which results in the absence of the precursors.

\begin{figure}[htbp]
\centering
\includegraphics[width=8cm]{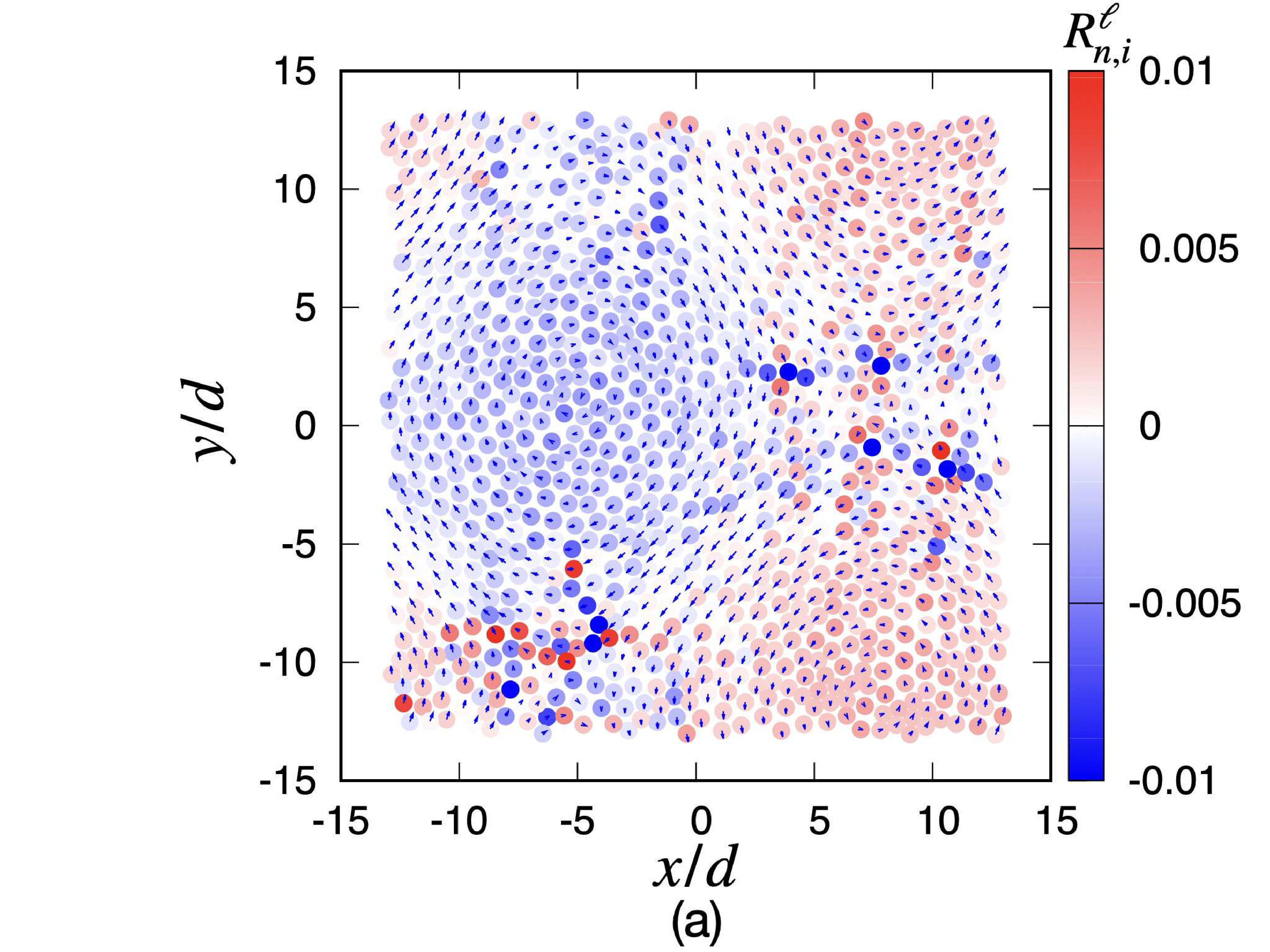}
\includegraphics[width=8cm]{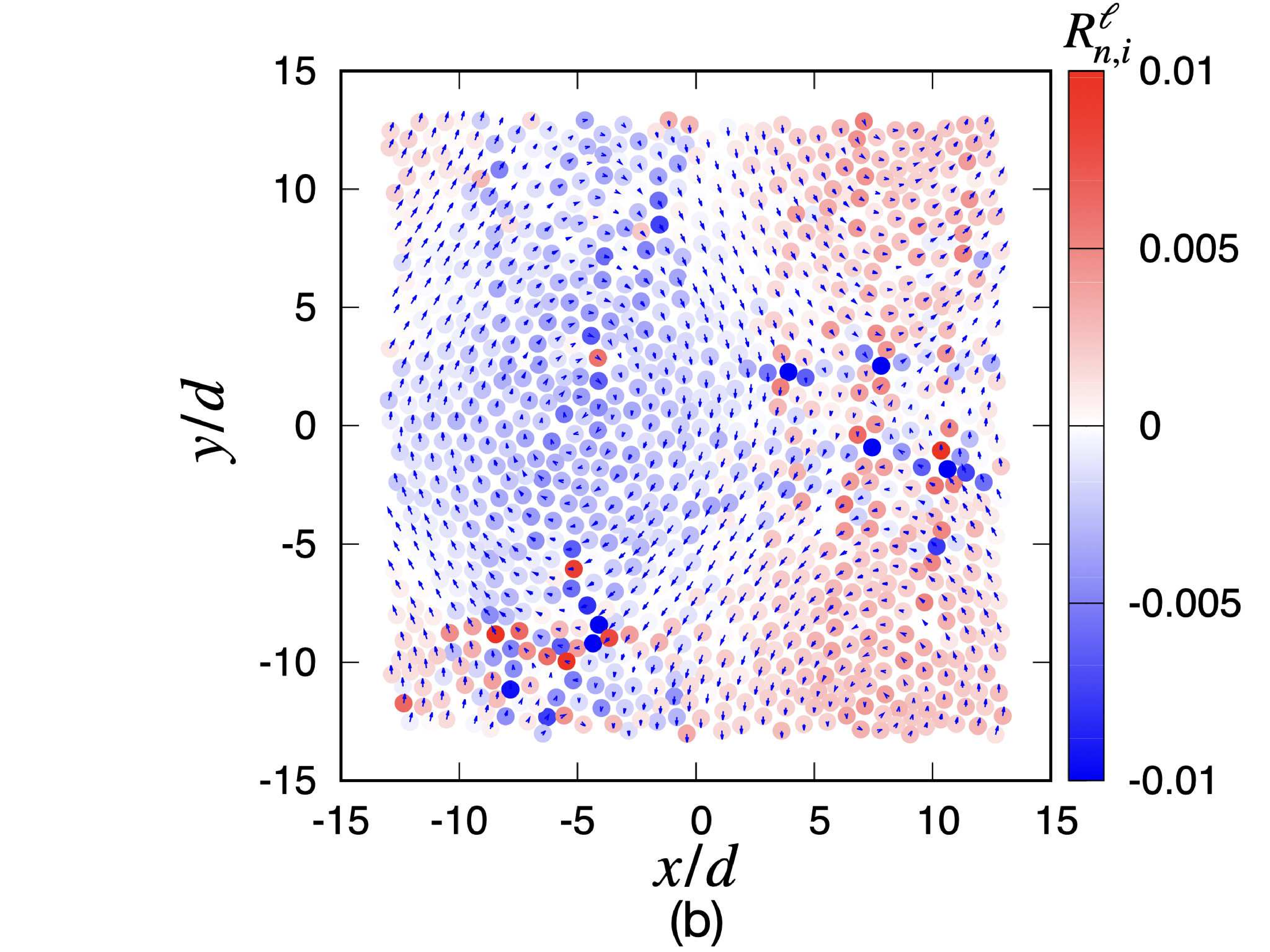}
\caption{
Plots of eigenvectors at (a) $\gamma_{c-}$  and at (b) $\gamma_{c+}$  corresponding to the smallest eigenvalue. 
For visualization, the magnitudes of the vectors are three times larger than their true values ($N=1024$).
}
\label{EiFieldDrop}
\end{figure}

Figure \ref{EiFieldDrop} shows a set of plots of the eigenvectors corresponding to the smallest eigenvalue at (a) $\gamma_{c-}$ and at (b)$\gamma_{c+}$, where $\gamma_{c+}$ is the strain immediately after the plastic event, 
and $\gamma_{c-}:=\gamma_{c+}-\Delta\gamma_{\rm Th}$ is the strain just before the event.
As shown in Fig.~\ref{EiFieldDrop}, changes in eigenvectors owing to the stress drop event can be observed.
Here, we find the existence of domains of grains of clockwise rotation and counter-clockwise rotation, and the collective motion of grains between two domains.
We may observe the excitation of the quadrupole-like mode, although its structure is not sufficiently clear.

\begin{figure}[htbp]
\centering
\includegraphics[width=8cm]{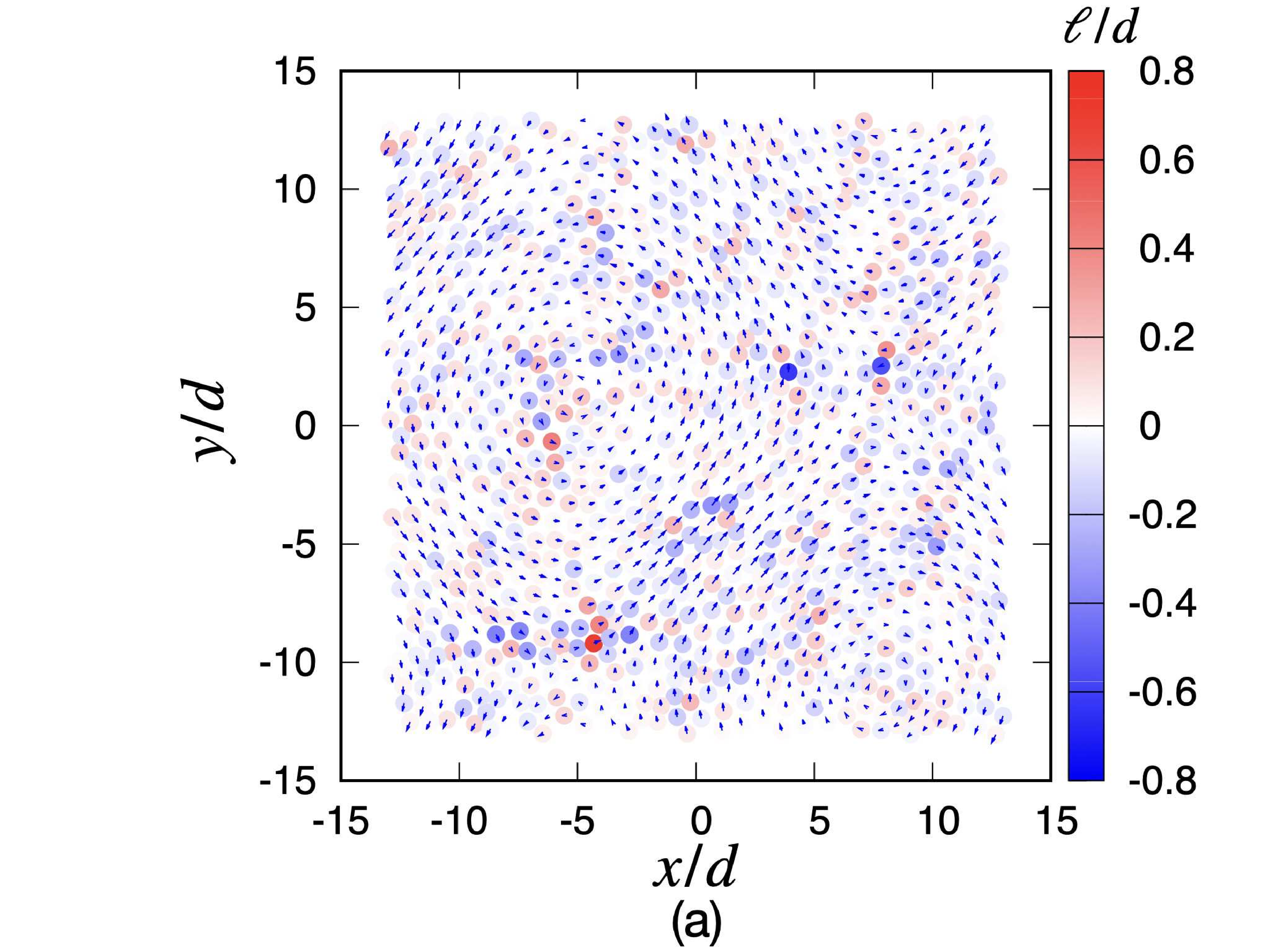}
\includegraphics[width=8cm]{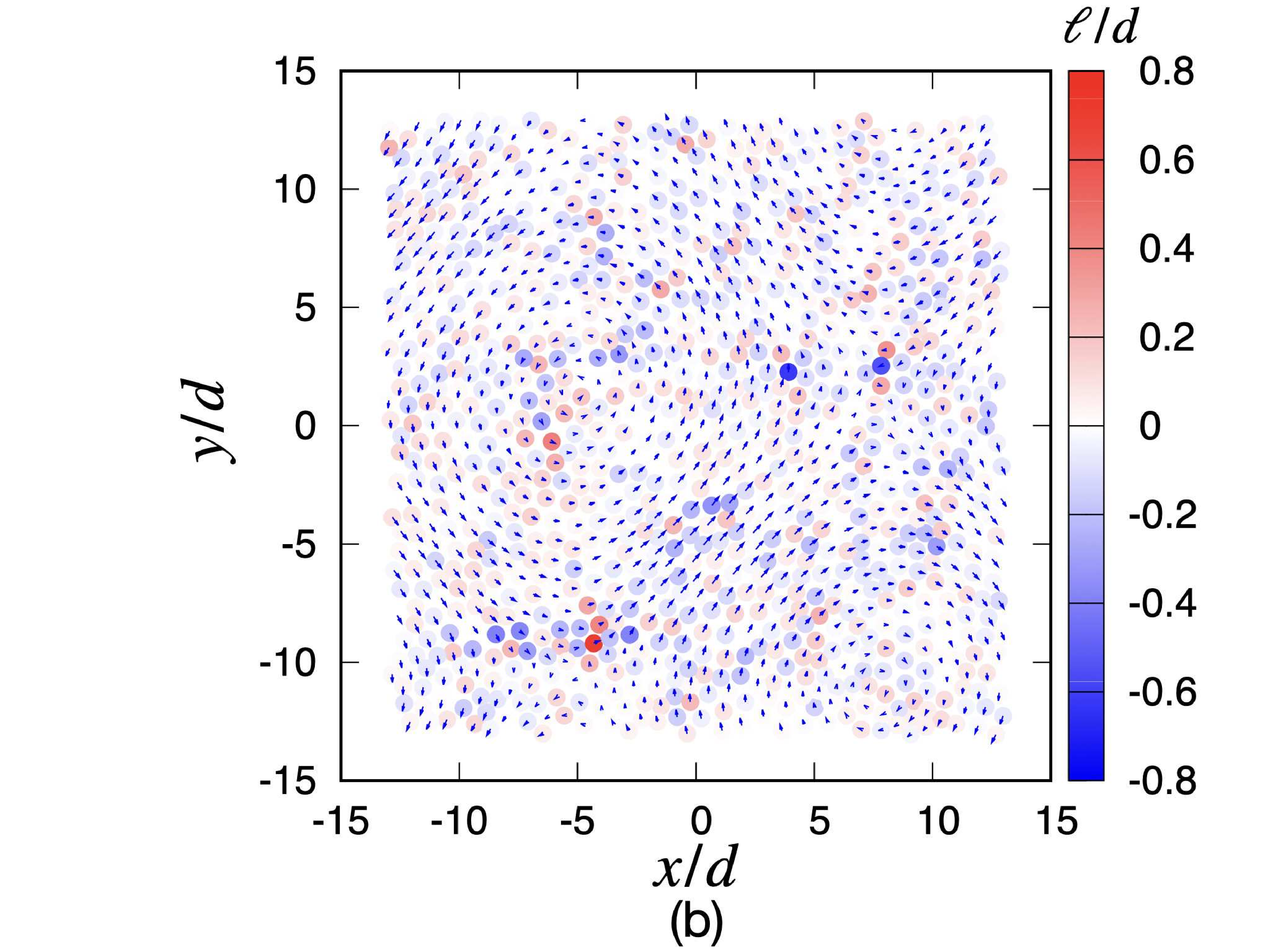}
\caption{
Plots of $\Ket{d\mathring q/d\gamma}$ at $\gamma=\gamma_{c+}$ for $N=1024$ and $\Delta\gamma_{\textrm{Th}}=1.0\times10^{-8}$, 
where (a) and (b) are based on the eigenvalue analysis and simulation, respectively. 
For visualization, we magnify  the magnitudes of the vectors with the factor $10.0$.
}
\label{NAFieldComp}
\end{figure}

Figure~\ref{NAFieldComp} is the comparison of $\Ket{d\mathring q/d\gamma}$ obtained by the eigenvalue analysis (a) with that by the simulation (b) at $\gamma=\gamma_{c+}$, where $\Ket{d\mathring q/d\gamma}$ in the simulation is evaluated by $(\mathring{\bm{q}}(\gamma_{c+}+\Delta\gamma_{\textrm{Th}}) - \mathring{\bm{q}}(\gamma_{c+}) )/\Delta\gamma_{\textrm{Th}}$ with $\Delta\gamma_{\textrm{Th}}=1.0\times10^{-8}$. 
It is obvious that the difference between the two figures is invisible, though the quadrupole-like structure cannot be clearly seen 
as in Fig.~\ref{EiFieldDrop}.
Nevertheless, we can find the collective motion of grains in both figures.

\begin{figure}[htbp]
\centering
\includegraphics[width=8cm]{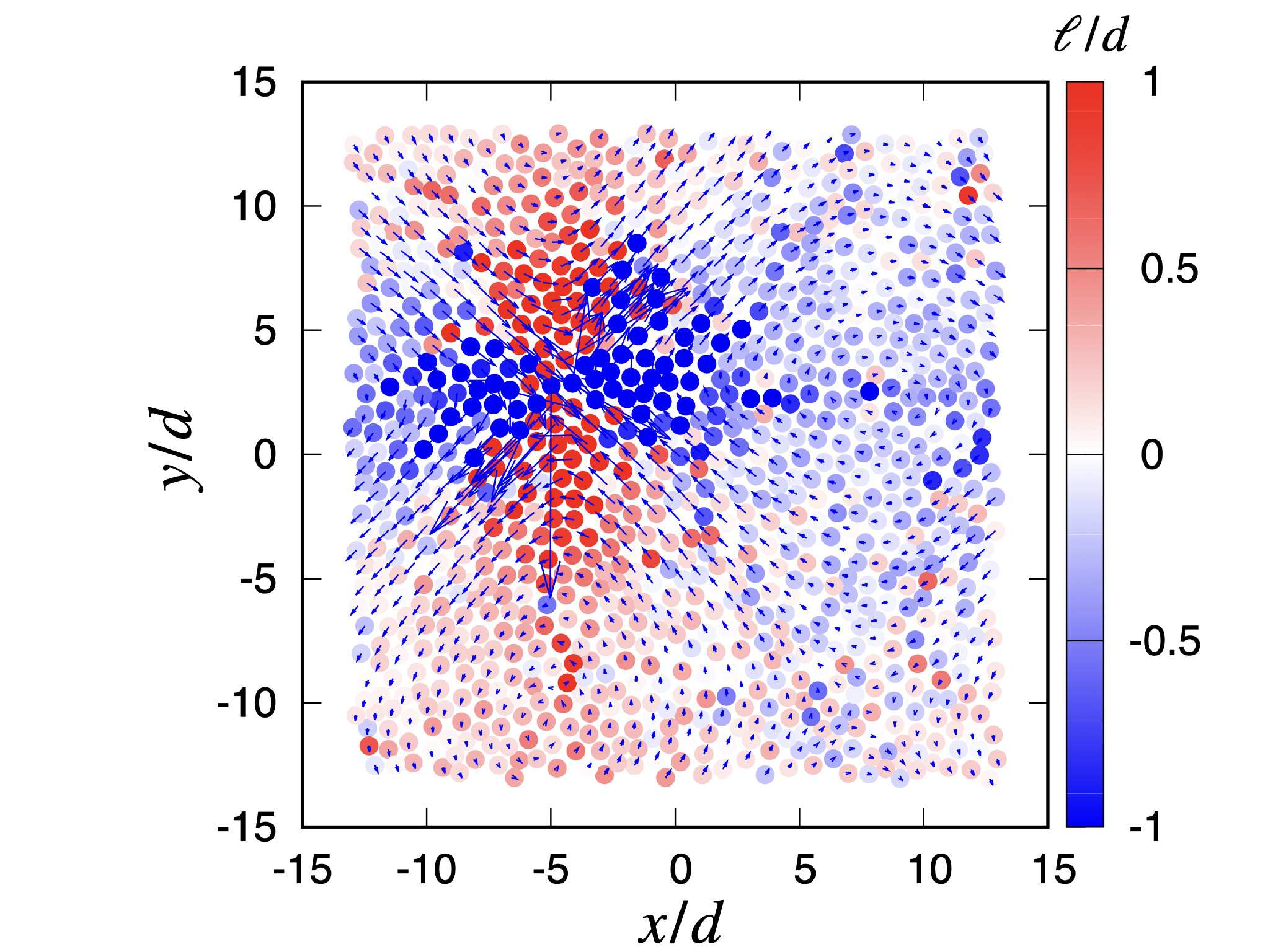}
\caption{
Plot of $\Delta\mathring{q}|_{c}$ in the simulation for $N=1024$ and $\Delta\gamma_{\textrm{Th}}=1.0\times10^{-8}$.
For visualization, we magnify $\Delta\mathring q|_{c}$ with the factor $1.0\times10^{3}$.
}
\label{NAFieldDropLrgN}
\end{figure}

Because we cannot use the eigenvalue analysis at the critical strain $\gamma_c$ for a plastic event, 
let us analyze the nonaffine displacement $\Delta \mathring{\bm{q}}$ between $\gamma_{c-}$ and $\gamma_{c+}$ caused by an avalanche using the simulation: 
\begin{align}
\Delta\mathring{\bm{q}}|_{c}
:=
\left[
\begin{matrix}
\left. \Delta\mathring{\bm{q}}_{1}\right|_{c} \\
\left. \Delta\mathring{\bm{q}}_{2} \right|_{c} \\
\vdots \\
\left. \Delta\mathring{\bm{q}}_{N} \right|_{c}
\end{matrix}
\right]
,
\end{align}
where
\begin{align}
\left.\Delta\mathring{\bm{q}}_{i}\right|_{c}:=
\left[
\begin{matrix}
r_{i}^{\textrm{FB},x}(\gamma_{c+}) - r_{i}^{\textrm{FB},x}(\gamma_{c-}) - \Delta\gamma r_{i}^{\textrm{FB},y}(\gamma_{c-}) \\
r_{i}^{\textrm{FB},y}(\gamma_{c+}) - r_{i}^{\textrm{FB},y}(\gamma_{c-}) \\
\ell_{i}^{\textrm{FB}}(\gamma_{c+}) - \ell_{i}^{\textrm{FB}}(\gamma_{c-})
\end{matrix}
\right]
.
\end{align}

Figure \ref{NAFieldDropLrgN} shows a plot of the nonaffine displacement $\Delta\mathring{\bm{q}}|_{c}$ around a yielding point based on the simulation for $N=1024$ at $\Delta\gamma_{\rm Th}=1.0\times10^{-8}$.  
This figure indicates that 
(i) grains move with rotations, which is one of the effects of mutual frictions between grains, and 
(ii) the existence of a quadrupole consisting of four domains of collective rotating grains exists.
In particular, the rotations of the grains are sharply changed on the boundary between the domains. 
Unfortunately, $\Delta\mathring{\bm{q}}|_{c}$ cannot be described by the eigenvalue analysis, because $\Delta\mathring{\bm{q}}|_{c}$ expresses the configuration change, which is unstable for a small change in $\gamma$.

\begin{figure}[htbp]
\centering
\includegraphics[width=7cm]{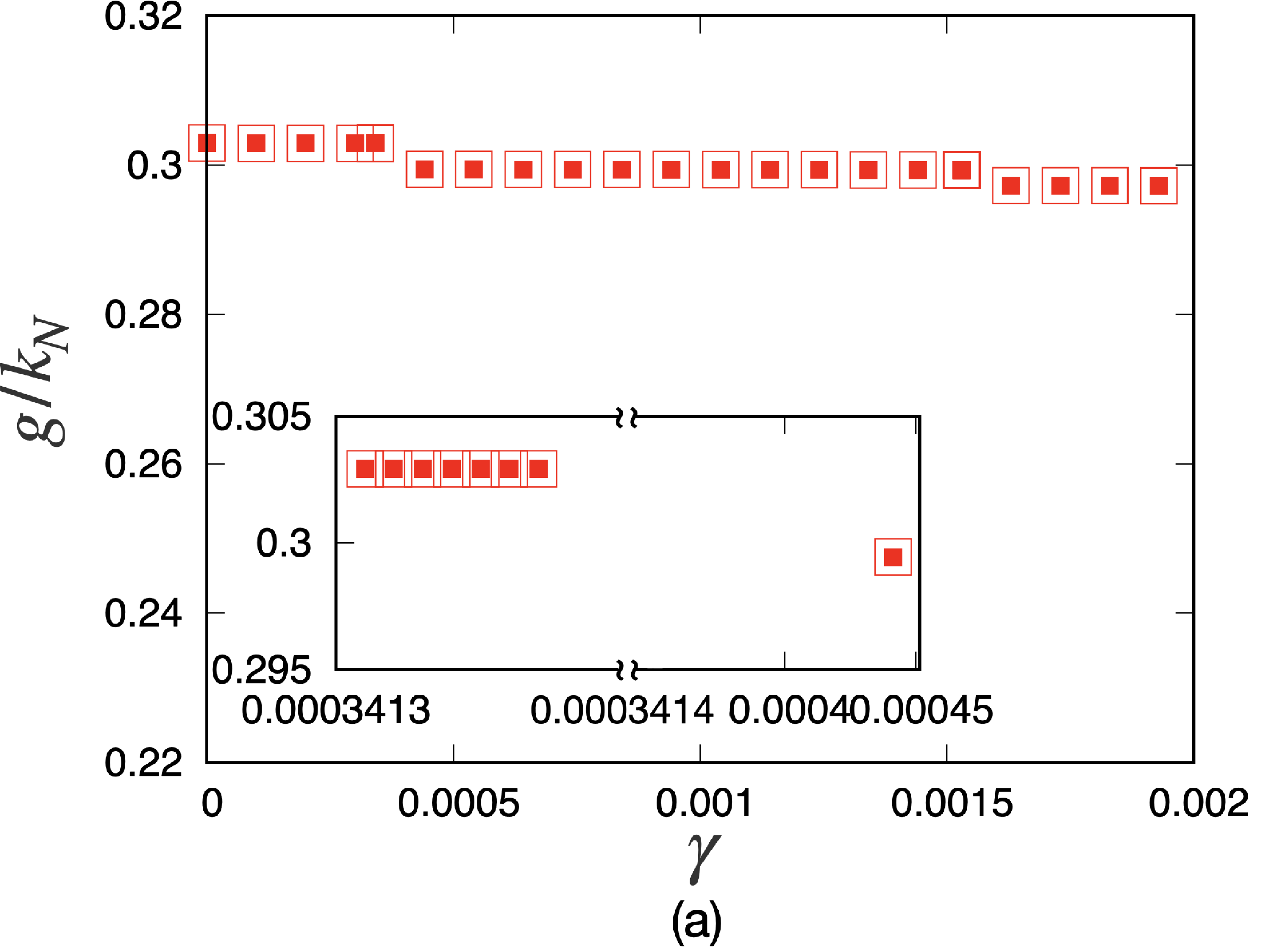}
\includegraphics[width=7cm]{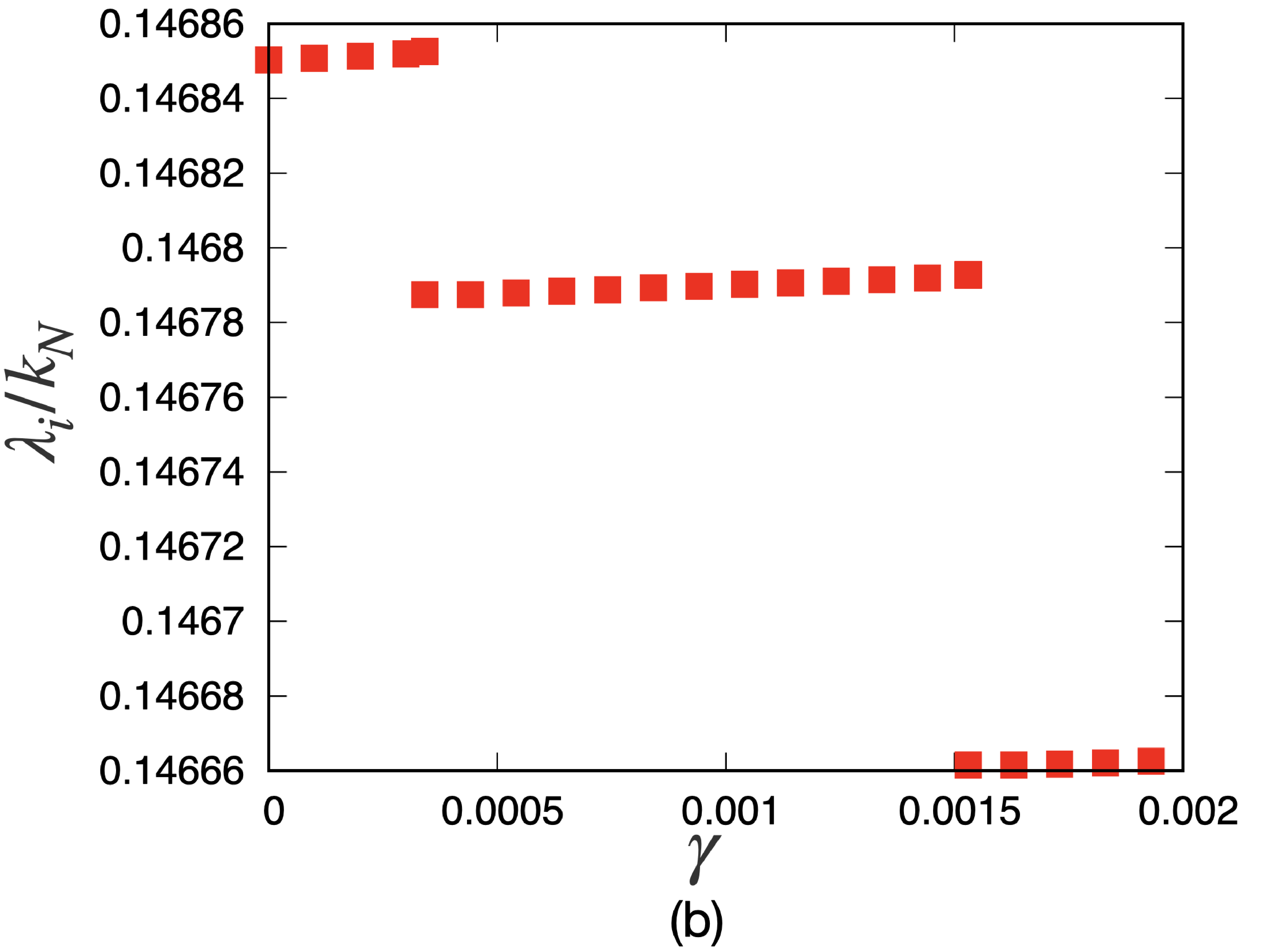}
\caption{
Plots of the shear modulus with (a) the theoretical evaluation (open symbols) and the simulation results (filled symbols) except for critical strain with the close-up of $g$ near a yielding point (inset), 
and (b) the smallest eigenvalue, except for zero modes 
against $\gamma$ for $0\leq\gamma\leq0.002$ for $N=128$. 
Note that the rigidity is not plotted at the yielding points, because it diverges there.
}
\label{Contact}
\end{figure}

Figures~\ref{Contact} (a) and (b) show plots of the rigidity and smallest eigenvalue from $\gamma=0$ to $0.002$, which includes two plastic events based on the one-sample calculation of the collection of grains with $N=128$.
One can find an almost perfect agreement of the rigidity between the eigenvalue analysis and simulation, except for the yielding points (see Fig.~\ref{Contact} (a)).
We find discontinuous changes in the smallest eigenvalue at the yielding point, where the rigidity changes discontinuously (see Fig. ~\ref{Contact} (b).
As expected, the magnitude of the discontinuous change in rigidity at the yielding point in Fig.~\ref{Contact} (a) for $\gamma\approx 0.001$ is smaller than that for a point of a stress drop for $\gamma\approx 0.341$, as shown in Fig.~\ref{Drop}.

\begin{figure}[htbp]
\centering
\includegraphics[width=8cm]{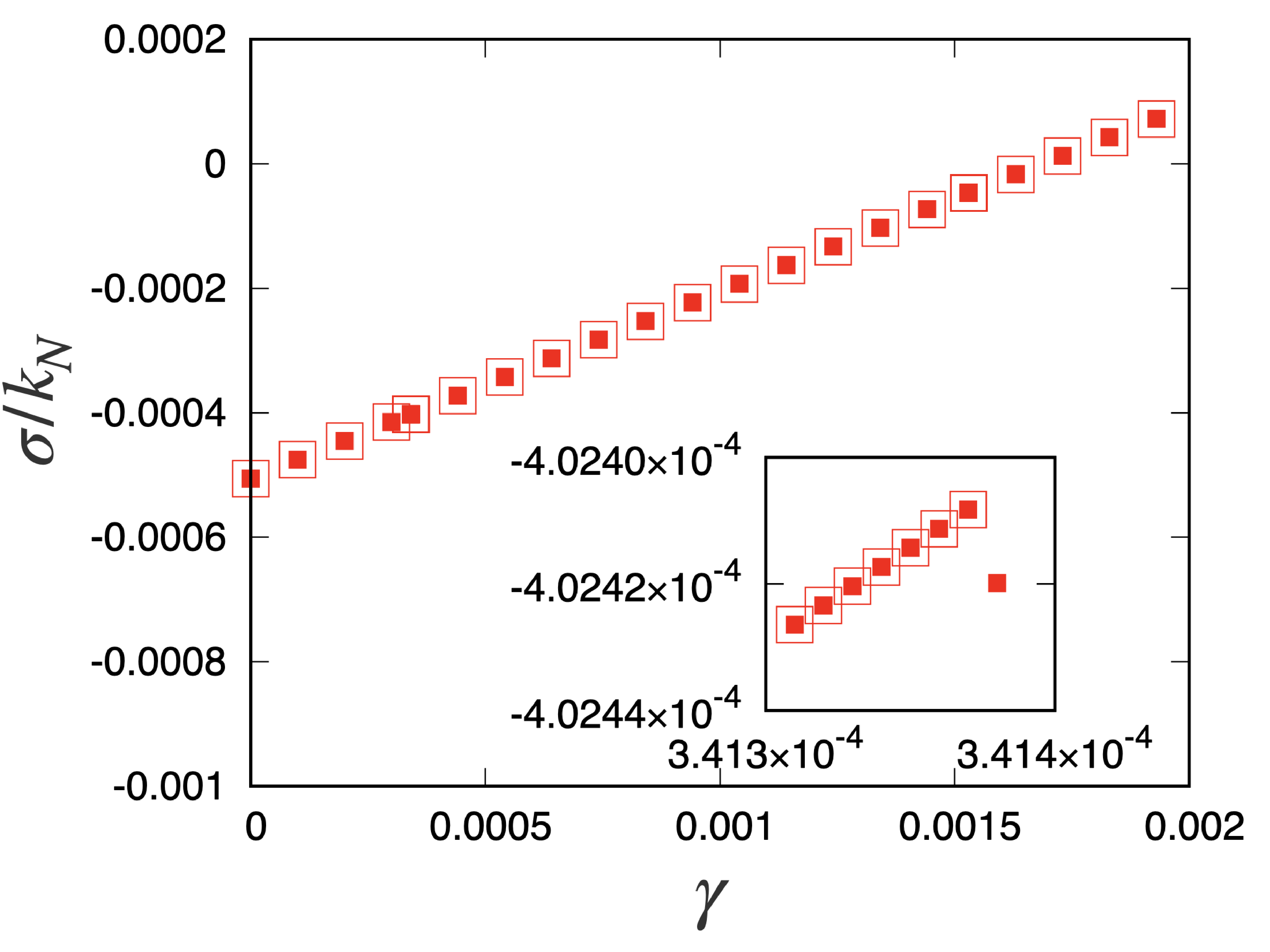}
\caption{
The plot of the stress-strain curve in the region $0\leq\gamma\leq0.002$ for $N=128$.
}
\label{StressContact}
\end{figure}

Figure~\ref{StressContact} shows the stress-strain curve corresponding to Fig.~\ref{Contact}.
It is difficult to find the plastic events in the main figure of Fig.~\ref{StressContact}, but we can find a small stress drop at this point if we use a close-up figure in the inset.
We verify the creation and annihilation of contacting pairs at the stress drop points.
The stress expression in Eq.~\eqref{theoretical_sigma} cannot be used at the yielding point; 
thus a disagreement exists between the eigenvalue analysis and simulation at the point in the inset of Fig. \ref{StressContact}.

\begin{figure}[htbp]
\centering
\includegraphics[width=8cm]{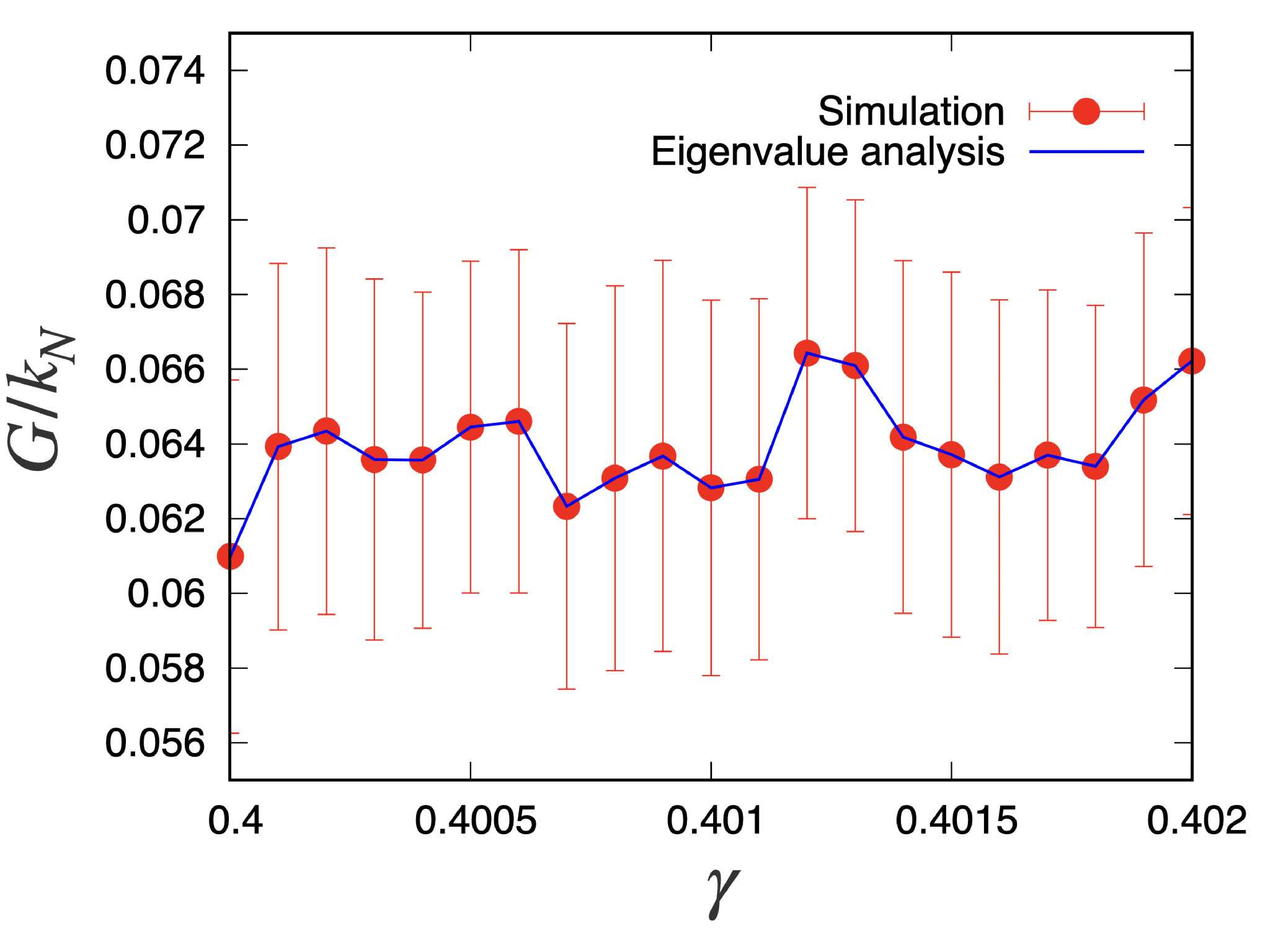}
\caption{
Plots of the rigidity $G$ based on the eigenvalue analysis (line) and on the simulation (filled symbols) with $\Delta\gamma_{\textrm{Th}}=1.0\times10^{-4}$.
We have used $30$ samples for $N=128$, where the error bars represent the standard deviations for $\gamma$. 
Note that we have omitted the data if stress-drop events occur.
}
\label{GContactEnsemble}
\end{figure}

Figure~\ref{GContactEnsemble} plots the rigidity of $G$ over 30 samples for $N=128$, 
where we have omitted the data if stress drop events take place.  
We verify that rigidity based on the eigenvalue analysis reproduces the results of the simulation.
Note that non-monotonic changes in $G$ originate from changes in the contact points and configuration of grains.

\section{Conclusion}\label{concluding_remarks}

In this paper, we have demonstrated that eigenvalue analysis of the Hessian matrix provides precise descriptions of the rigidity and stress of dispersed frictional grains 
in which the contact force is described by the harmonic potential, in spite of stress-drop events, such as stress avalanches. 
However, our model does not contain any slip processes between contacting grains.
This success is a natural extension of the previous studies on frictionless grains~\cite{Maloney04,Maloney06} to frictional grains and of our previous study on the linear response regime~\cite{Ishima2022} to the nonlinear response regime.
Two remarkable features of the contacting model are described by the harmonic potential.
First, the tangential contact force in this model is no longer a history-dependent one.
This leads to the significant simplification of the theoretical analysis. 
Second, unlike the naive expectation, the eigenvalues in our model do not indicate any precursors for the stress-drop events.
In essence, stress-drop events take place suddenly by releasing contact points.

Some future tasks that need to be addressed are as follows.
First, we need to consider the effect of slips, which causes a significant difference from our model because history-dependent contacts play important roles in the presence of slip events.
Second, we must extend our analysis to nonlinear interacting models, such as the Hertzian contact model in a three-dimensional space.
We plan on working on these points in the future.

\section{ACKNOWLEDGMENTS}
The authors thank N. Oyama for the fruitful discussions and useful comments.
This work was partially supported by a Grant-in-Aid from the MEXT for Scientific Research (Grant No. JP22K03459, JP21H01006, and JP19K03670) and a Grant-in-Aid from the Japan Society for Promotion of Science JSPS Research Fellow (Grant No. JP20J20292).

\appendix

\begin{widetext}


\section{Absence of the second term on RHS of Eq.~\eqref{xiT}}\label{App:Absence_of_2nd}

\begin{figure}[htbp]
\centering
\includegraphics[width=8cm]{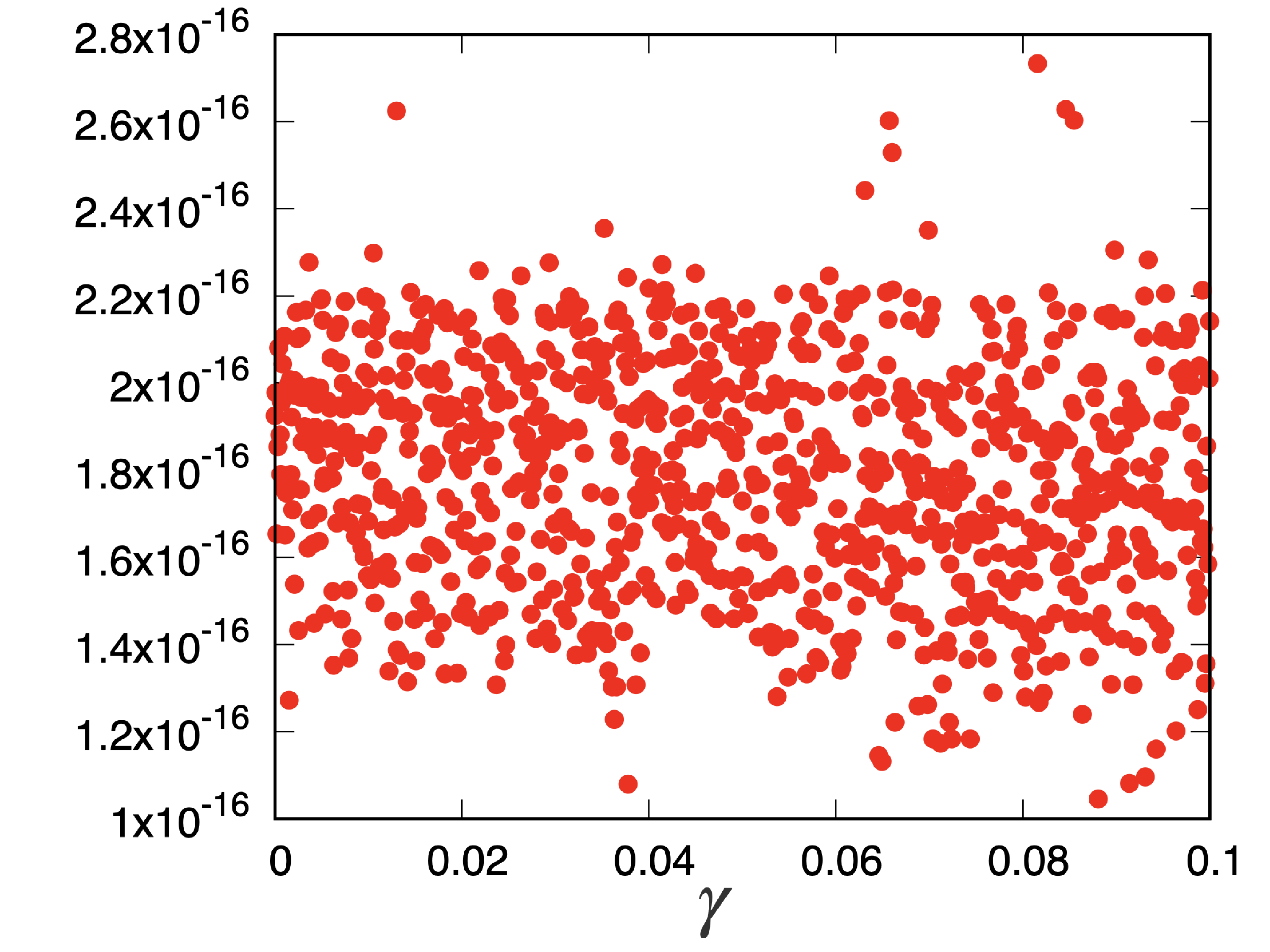}
\caption{
Plot of the maximum $\textrm{Err}$ against $\gamma$.
}
\label{err}
\end{figure}

Thus far, we could not prove that the second term on RHS of Eq.~\eqref{xiT} can be regarded as zero with numerical accuracy, 
but we verify that this term is zero, at least, in the numerical simulation of harmonic systems as follows:
Let us calculate the ratio of the second term in Eq.~\eqref{xiT} to the first term using
\begin{align}
\textrm{Err}:=\frac{\left| \left[\left( \int_{C_{ij}}dt\bm{v}_{T,ij}\right)\cdot\bm{n}_{ij}\right]\bm{n}_{ij} \right|}{\left|\int_{C_{ij}}dt\bm{v}_{T,ij}\right|} .
\end{align}
Figure \ref{err} shows the plots of the largest $\textrm{Err}$ in contacting pairs against $\gamma$, which indicates $|{\rm Err}|<3\times 10^{-16}$.
As our calculation is based on double precision, which has only 16 significant digits, $\textrm{Err}$ can be regarded as zero.

\section{The behavior of the smallest eigenvalue near stress-drop points}\label{app:gamma_Th}

In this appendix, we provide an in-depth explain the behavior of the smallest eigenvalue in the vicinity of the stress-drop points in detail. 
We adopt the following protocol to reduce the step strain small in the vicinity of the stress-drop point.
We adopt $\Delta\gamma_{\rm in}=10^{-4}$ in the appendix.
We use $\Delta\gamma=\Delta\gamma_{\rm in}$ if there is no plastic event during the strain increment $\Delta\gamma$.
If we find a stress drop during the strain from $\gamma$ to $\gamma+\Delta\gamma$, we restore the system to the state $\gamma$,
and examine $\gamma+0.1\Delta\gamma_{\rm in}$.
If we do not find any stress drop, we further add the strain with $\Delta\gamma_{\rm in}$; if we still have a stress drop, we repeat the procedure of restoring and adding strain $0.01\Delta\gamma_{\rm in}$.
This protocol is repeated to detect stress drop events until we reach  $\Delta\gamma<\Delta\gamma_{\rm Th}$.          
In this appendix, we illustrate how the results depend on the choice of $\Delta\gamma_{\rm Th}$, where the smallest value of $\Delta\gamma_{\rm Th}$ is $10^{-10}$.

Figure~\ref{SxySmlEE} presents the stress-strain curves obtained using this protocol. 
The upper branch in Fig.~\ref{SxySmlEE} represents the shear stress below the stress drop, and the lower branch represents the shear stress above the stress drop.
The smallest $\gamma$ in the lower branch and the largest $\gamma$ in the upper branch strongly depend on $\Delta\gamma_{\rm Th}$.
As shown in Fig.~\ref{CriticalGamma}, the stress drop takes place at $\gamma\approx 0.01330$ for $\Delta\gamma_{\rm Th}=10^{-4}$, whereas the critical strain $\gamma_c$ for the stress drop approaches 0.013334 as $\Delta\gamma_{\rm Th}$ decreases, where $\gamma_c$ is 0.013334 for $\Delta\gamma_{\rm Th}\le 10^{-6}$.

\begin{figure}[htbp]
\centering
\includegraphics[width=8cm]{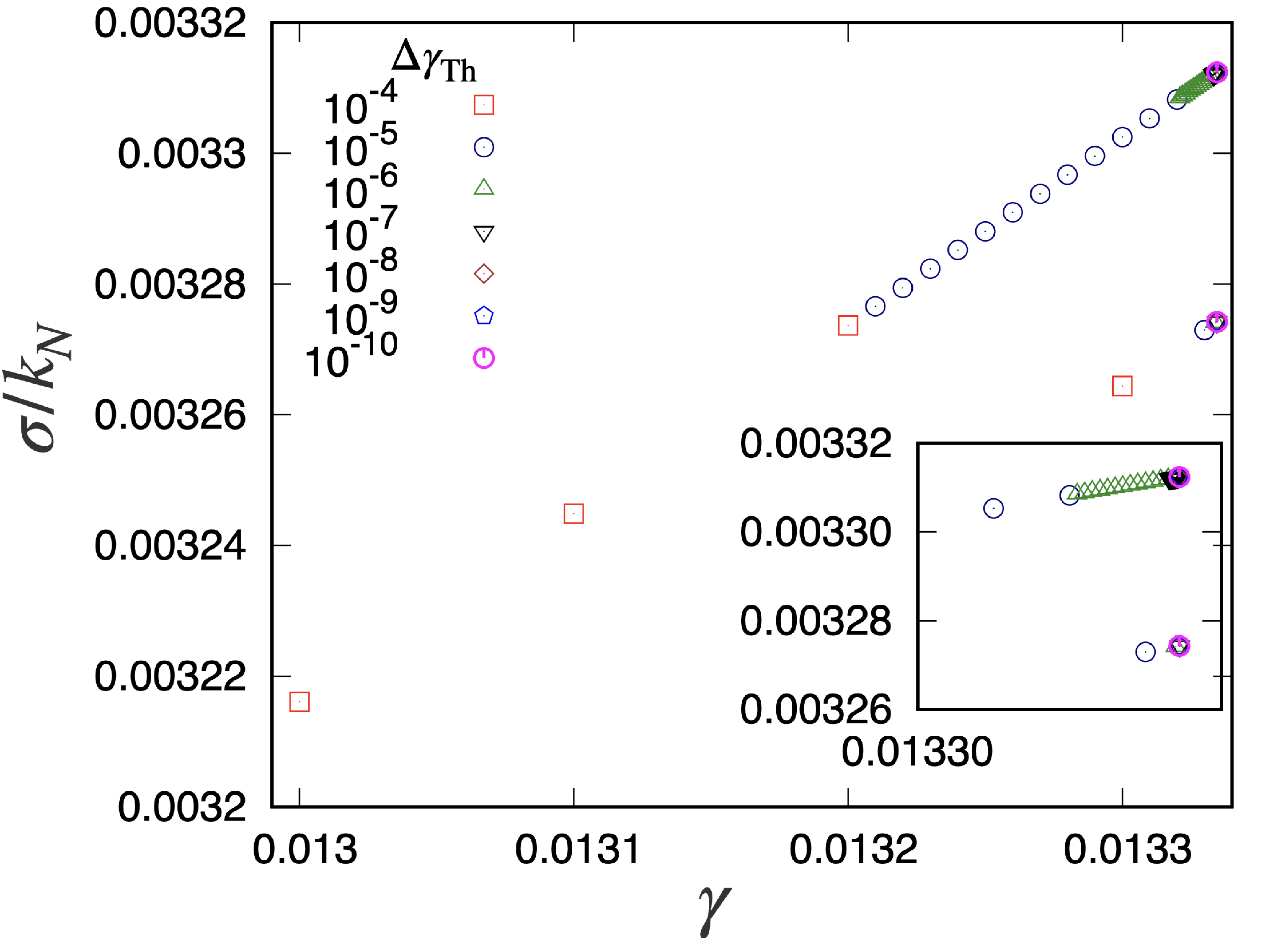}
\caption{
The stress-strain curve for $N=128$, which are the stress drop points for various $\Delta\gamma_{\rm Th}$ (from $\Delta\gamma_{\rm Th}=10^{-4}$ and $\Delta\gamma_{\rm Th}=10^{-10}$). 
}
\label{SxySmlEE}
\end{figure}

\begin{figure}[htbp]
\centering
\includegraphics[width=8cm]{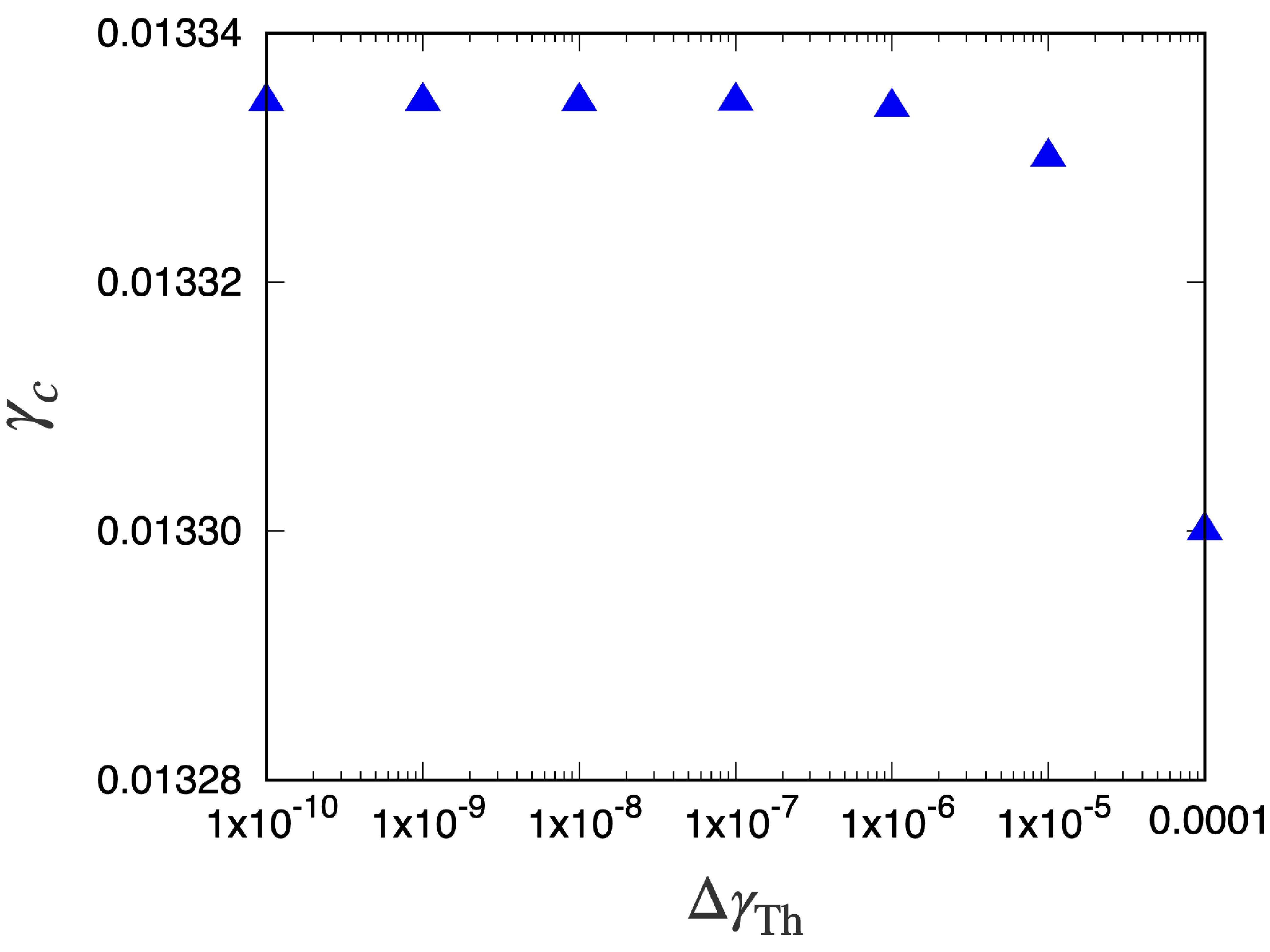}
\caption{
Plot of the critical strain $\gamma_{c}$ against $\Delta\gamma_{\rm Th}$ for $N=128$.
}
\label{CriticalGamma}
\end{figure}

Figure~\ref{EiSmlEE} plots the behavior of the smallest eigenvalue against $\gamma$ corresponding to Fig.~\ref{SxySmlEE} for
$\Delta\gamma_{\rm Th}=10^{-10}$ immediately below the stress drop point, 
where the symbols correspond to the analysis for the corresponding $\Delta\gamma$ as in Fig.~\ref{SxySmlEE}.
We have confirmed that there is no precursor of the eigenvalues below $\gamma_c$ as observed in Hertzian and Lennard-Jones systems~\cite{Maloney04,Maloney06,Manning11}. 
Thus, the harmonic system does not have any precursors in the behavior of the smallest eigenvalue.

\begin{figure}[htbp]
\centering
\includegraphics[width=8cm]{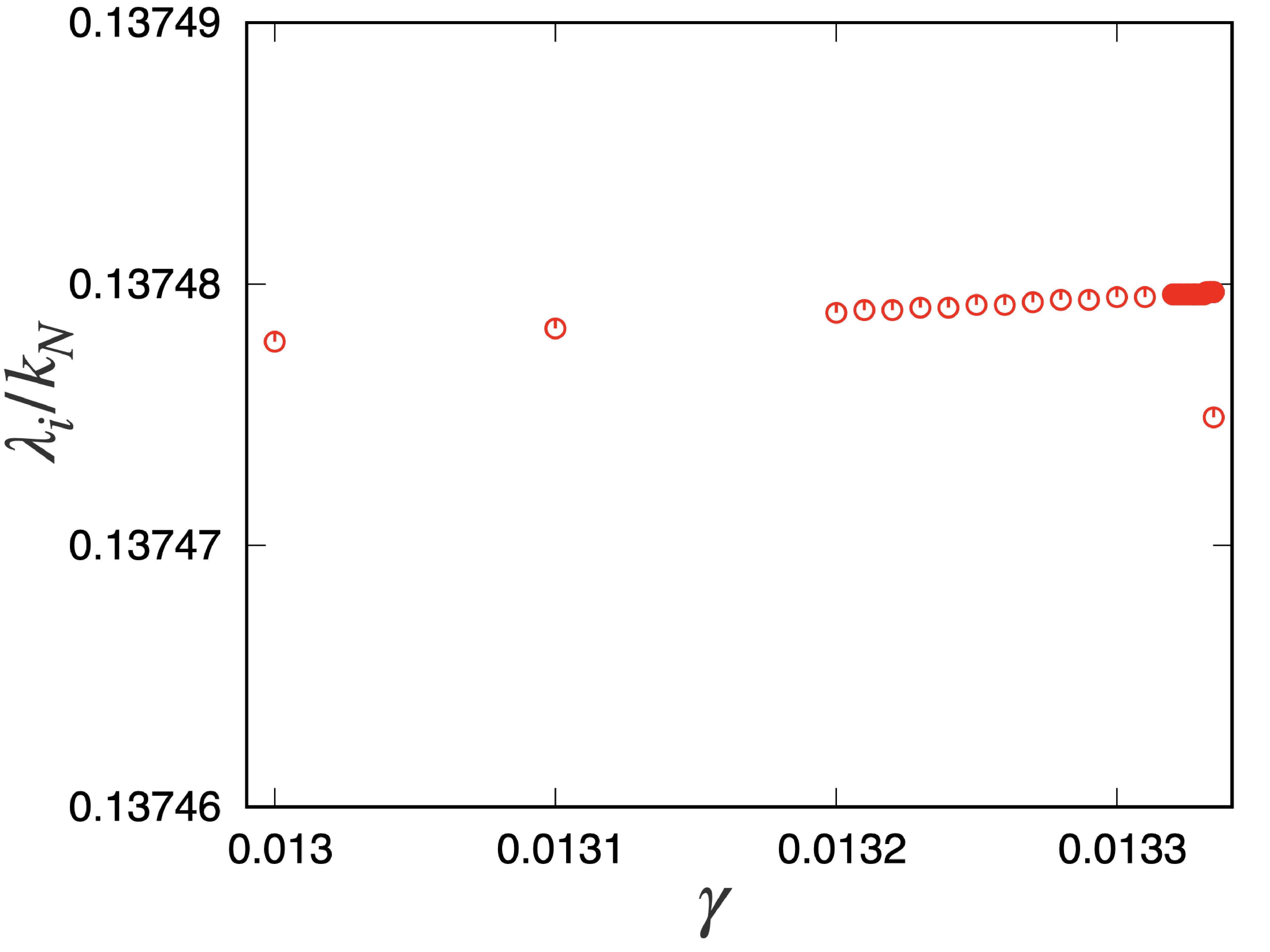}
\caption{
The plot of the smallest non-zero eigenvalue in the vicinity of $\gamma_c$ for $\Delta\gamma_{\rm Th}=1.0\times10^{-10}$.
}
\label{EiSmlEE}
\end{figure}

\section{Some properties of the Hessian matrix in a harmonic system \label{AppHessian}}
In this appendix, we briefly summarize the  properties of the Hessian matrix of the harmonic contact model. 
In Sec. \ref{Hessian=Jacobian}, we explicitly express the elements of the Hessian matrix in this model.
In Sec. \ref{app:eff_boundary}, we demonstrate that the symmetry of the Hessian matrix still holds even under the Lees-Edwards boundary condition.

\subsection{The explicit expression for the Hessian matrix}\label{Hessian=Jacobian}

In this appendix, we present an explicit expression for the Hessian matrix.
To this end, we return to the effective potential in Eq.~\eqref{potential}.
It is straightforward to obtain
\begin{align}
\frac{\partial^{2} \delta\bm{r}_{ij,\perp}^{2}}{\partial x_{i}^{2}}&=2 - 2(n_{ij}^{x})^2
\label{xx}
,\\
\frac{\partial^{2} \delta\bm{r}_{ij,\perp}^{2}}{\partial x_{i}\partial y_{i}}&=-2n_{ij}^{x}n_{ij}^{y}
,\\
\frac{\partial^{2} \delta\bm{r}_{ij,\perp}^{2}}{\partial x_{i}\partial \ell_{i}}&=
2n_{ij}^{y} 
,\\
\frac{\partial^{2} \delta\bm{r}_{ij,\perp}^{2}}{\partial y_{i}\partial x_{i}}&=
\frac{\partial^{2} \delta\bm{r}_{ij,\perp}^{2}}{\partial x_{i}\partial y_{i}}
,\\
\frac{\partial^{2} \delta\bm{r}_{ij,\perp}^{2}}{\partial y_{i}^{2}}&=2 - 2(n_{ij}^{y})^2
,\\
\frac{\partial^{2} \delta\bm{r}_{ij,\perp}^{2}}{\partial y_{i}\partial \ell_{i}}&=
2n_{ij}^{x} 
,\\
\frac{\partial^{2} \delta\bm{r}_{ij,\perp}^{2}}{\partial \ell_{i}\partial x_{i}}&=
\frac{\partial^{2} \delta\bm{r}_{ij,\perp}^{2}}{\partial x_{i}\partial \ell_{i}}
,\\
\frac{\partial^{2} \delta\bm{r}_{ij,\perp}^{2}}{\partial \ell_{i}\partial y_{i}}&=
\frac{\partial^{2} \delta\bm{r}_{ij,\perp}^{2}}{\partial y_{i}\partial \ell_{i}}
,\\
\frac{\partial^{2} \delta\bm{r}_{ij,\perp}^{2}}{\partial \ell_{i}^{2}}&=
2 .
\label{ll}
\end{align}

Thus, we obtain
\begin{align}
\mH_{ij}^{xx}&
=\frac{\partial^{2} \delta e_{ij}}{\partial x_{i}\partial x_{j}} 
=- \frac{\partial^{2} \delta e_{ij}}{\partial x_{i}^{2}}
=
-k_{N}+k_{N}\left[1+\frac{\xi_{N,ij}}{|\bm{r}_{ij}|}\right](n_{ij}^{y})^{2}
- k_{T}(n_{ij}^{y})^{2}
\label{xxH}
,\\
\mH_{ij}^{xy}&
= \frac{\partial^{2} \delta e_{ij}}{\partial x_{i}\partial y_{j}} 
=- \frac{\partial^{2} \delta e_{ij}}{\partial x_{i}\partial y_{i}}
=
- k_{N}\left[1+\frac{\xi_{N,ij}}{|\bm{r}_{ij}|}\right]n_{ij}^{x}n_{ij}^{y}
+ k_{T} n_{ij}^{x}n_{ij}^{y}
,\\
\mH_{ij}^{x\ell}&
= \frac{\partial^{2} \delta e_{ij}}{\partial x_{i}\partial \ell_{j}} 
=\frac{\partial^{2} \delta e_{ij}}{\partial x_{i}\partial \ell_{i}}
=k_{T}n_{ij}^{y}
,\\
\mH_{ij}^{yy}&
= \frac{\partial^{2} \delta e_{ij}}{\partial y_{i}\partial y_{j}} 
=- \frac{\partial^{2} \delta e_{ij}}{\partial y_{i}^{2}}
=
- k_{N} + k_{N}\left[1+\frac{\xi_{N,ij}}{|\bm{r}_{ij}|}\right](n_{ij}^{x})^{2}
- k_{T}(n_{ij}^{x})^{2}
,\\
\mH_{ij}^{y\ell}&
= \frac{\partial^{2} \delta e_{ij}}{\partial y_{i}\partial \ell_{j}} 
=\frac{\partial^{2} \delta e_{ij}}{\partial y_{i}\partial \ell_{i}}
=-k_{T}n_{ij}^{x}
,\\
\mH_{ij}^{\ell\ell}&
= \frac{\partial^{2} \delta e_{ij}}{\partial \ell_{i}\partial \ell_{j}} 
=\frac{\partial^{2} \delta e_{ij}}{\partial \ell_{i}^{2}}
=k_{T} 
\end{align}
for $i\neq j$ and 
\begin{align}
\mH_{ii}^{xx}&
= \sum_{j\neq i}\frac{\partial^{2} \delta e_{ij}}{\partial x_{i}^{2}}
= -\sum_{j\neq i}\left[
-k_{N}+k_{N}\left[1+\frac{\xi_{N,ij}}{|\bm{r}_{ij}|}\right](n_{ij}^{y})^{2}
- k_{T}(n_{ij}^{y})^{2}
\right]
,\\
\mH_{ii}^{xy}&
=\sum_{j\neq i} \frac{\partial^{2} \delta e_{ij}}{\partial x_{i}\partial y_{i}}
=
-\sum_{j\neq i}\left[
- k_{N}\left[1+\frac{\xi_{N,ij}}{|\bm{r}_{ij}|}\right]n_{ij}^{x}n_{ij}^{y}
+ k_{T} n_{ij}^{x}n_{ij}^{y}
\right]
,\\
\mH_{ii}^{x\ell}&
=\sum_{j\neq i} \frac{\partial^{2} \delta e_{ij}}{\partial x_{i}\partial \ell_{i}}
=\sum_{j\neq i} k_{T}n_{ij}^{y}
,\\
\mH_{ii}^{yy}&
= \sum_{j\neq i} \frac{\partial^{2} \delta e_{ij}}{\partial y_{i}^{2}}
=\sum_{j\neq i}\left[
- k_{N} + k_{N}\left[1+\frac{\xi_{N,ij}}{|\bm{r}_{ij}|}\right](n_{ij}^{x})^{2}
- k_{T}(n_{ij}^{x})^{2}
\right]
,\\
\mH_{ii}^{y\ell}&
=\sum_{j\neq i}\frac{\partial^{2} \delta e_{ij}}{\partial y_{i}\partial \ell_{i}}
=-\sum_{j\neq i}k_{T}n_{ij}^{x}
,\\
\mH_{ii}^{\ell\ell}&
=\sum_{j\neq i}\frac{\partial^{2} \delta e_{ij}}{\partial \ell_{i}^{2}}
=\sum_{j\neq i}k_{T}
\label{llH}
.
\end{align}

\subsection{Effect of the boundary condition to the Hessian matrix}\label{app:eff_boundary}

In this appendix, we explain the influence of strain and the boundary condition on the Hessian matrix in detail to determine 
whether the symmetry of the Hessian matrix is still maintained, even if we consider a system with non-zero strain.

First, let us consider the case in which particle $i$ interacts with the particle $j$ through a mirror image in $x$-direction, as shown in Fig.~\ref{xBdry}.
In this case, $\bm{r}_{ij}$ is given by
\begin{align}
\bm{r}_{ij}:=\bm{r}_{i}-\bm{r}_{j}-L\bm{e}_{x} ,
\end{align}
where $\bm{e}_{x}:=(1,0)^{\textrm{T}}$.
Similarly, $\bm{r}_{ji}$ is given by
\begin{align}
\bm{r}_{ji}:=\bm{r}_{j}-\bm{r}_{i}+L\bm{e}_{x} .
\end{align}
Thus, $\bm{r}_{ij}$ satisfies
\begin{align}
\bm{r}_{ji}=-\bm{r}_{ij} .
\end{align}
Then, we obtain
\begin{align}
\xi_{N,ij}
:&=\frac{d_{i}+d_{j}}{2}-|\bm{r}_{ij}| \nonumber \\
&=\frac{d_{i}+d_{j}}{2}-|\bm{r}_{ji}| \nonumber \\
&=\xi_{N,ji} .
\label{XiNBdryX}
\end{align}
Thus, the result is independent of the strain, and the symmetry of the Hessian is still valid in this case.

\begin{figure}[htbp]
\centering
\includegraphics[width=8cm]{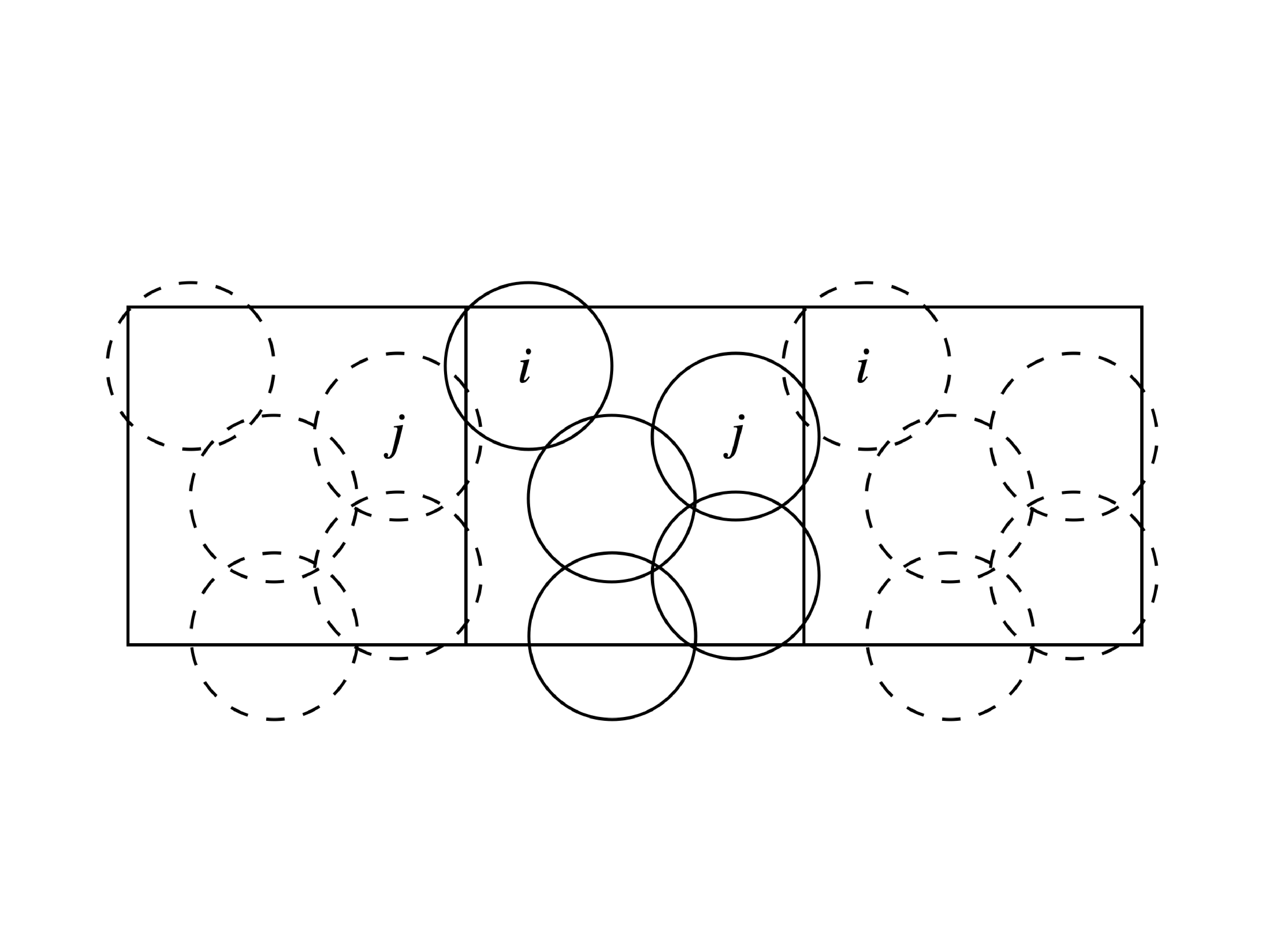}
\caption{
A schematic of the case where the particle $i$ interacts with the particle $j$ through the mirror image in $x$-direction.
}
\label{xBdry}
\end{figure}

Next, let us consider the case in which the particle $i$ interacts with the particle $j$ through a mirror image in the $y$-direction (see Fig.~\ref{yBdry}).
In this case, 
$\bm{r}_{ij}$ is given by
\begin{align}
\bm{r}_{ij}:=\bm{r}_{i}-\bm{r}_{j}+L\bm{e}_{y}+\gamma L\bm{e}_{x} ,
\label{rijY}
\end{align}
where $\bm{e}_{y}:=(0,1)^{\textrm{T}}$.
Similarly, $\bm{r}_{ji}$ is given by
\begin{align}
\bm{r}_{ji}:=\bm{r}_{j}-\bm{r}_{i}-L\bm{e}_{y}-\gamma L\bm{e}_{x} .
\label{rjiY}
\end{align}
Since the relation
\begin{align}
\bm{r}_{ji}=-\bm{r}_{ij},
\end{align}
we obtain
\begin{align}
\xi_{N,ij}
:&=\frac{d_{i}+d_{j}}{2}-|\bm{r}_{ij}| \nonumber \\
&=\frac{d_{i}+d_{j}}{2}-|\bm{r}_{ji}| \nonumber \\
&=\xi_{N,ji} .
\label{XiNBdryY}
\end{align}
Thus, $\xi_{N,ij}$ and $\xi_{N,ji}$ depend on $\gamma$ in the same way.
With the aid of Eqs.~\eqref{rijY},\eqref{rjiY} and \eqref{XiNBdryY}, the Hessian matrix depends on $\gamma$ if the particle interacts with another particle through the mirror image in $y$-direction, although the symmetry of the Hessian is still maintained.  

\begin{figure}[htbp]
\centering
\includegraphics[width=8cm]{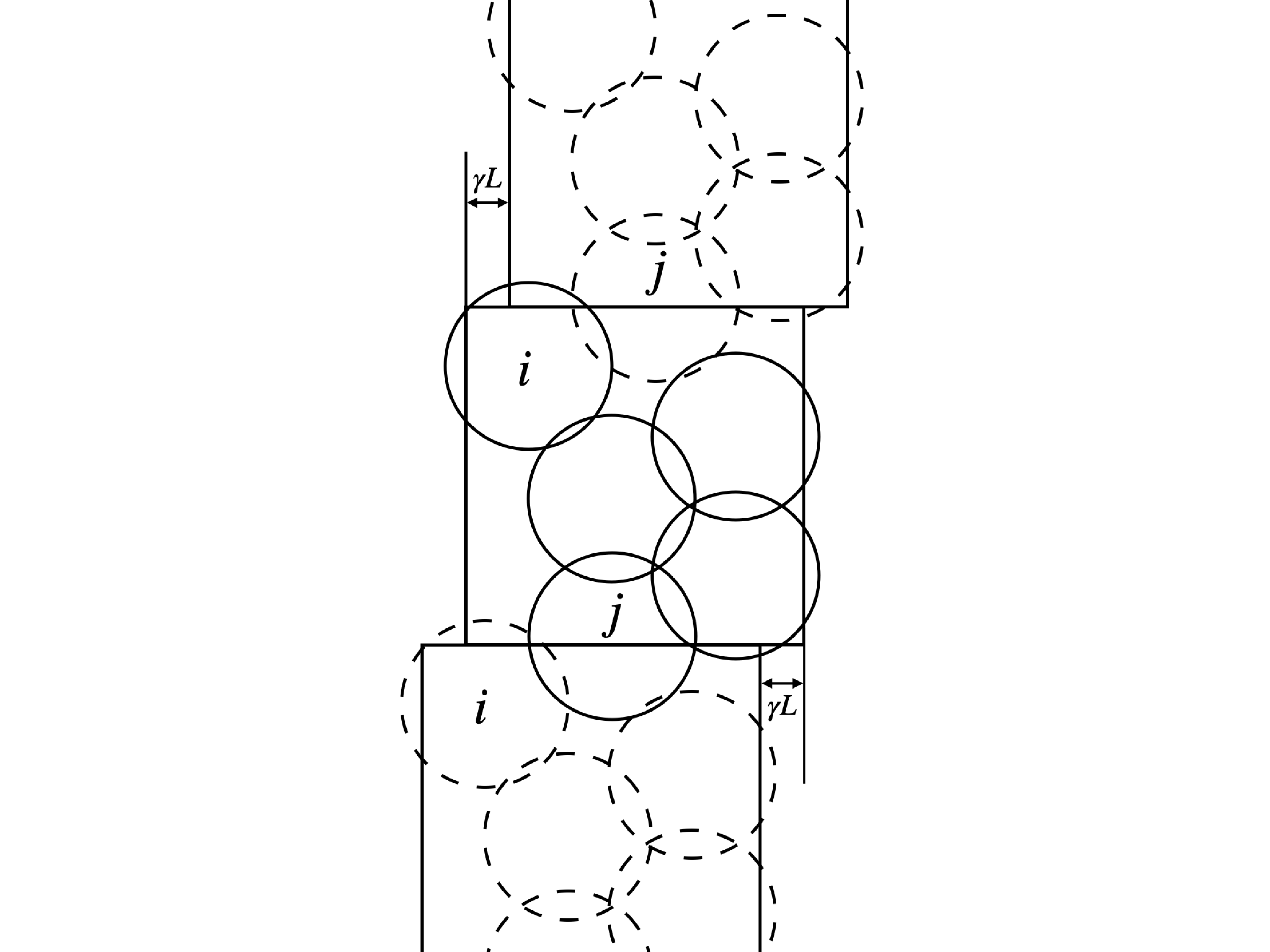}
\caption{
A schematic of the case that the particle $i$ interacts with the particle $j$ through the mirror image in $y$-direction.
}
\label{yBdry}
\end{figure}


\section{Some properties of the Jacobian matrix in a harmonic system and its equivalency to the Hessian matrix~\label{AppHarmonic}}

In this appendix, we briefly summarize the properties of the Jacobian matrix for the harmonic contact model that was previously used in the description of frictional grains~\cite{Chattoraj19,Chattoraj19E, Charan20, Ishima2022}.
In Sec.\ref{AppHkn0}, we present explicit forms of the diagonal and non-diagonal blocks of the Jacobian matrix. 
In Sec.\ref{AppHkn1}, we present the derivation of the Jacobian for the harmonic contact model.
In Sec. \ref{AppHkn2}, we explicitly write the elements of the Jacobian matrix in the model.

\subsection{Jacobian block elements\label{AppHkn0}}
Let us write a $3\times3$ sub-matrix $\mJ_{ij}$, which is the $(ij)$ block element of the Jacobian obtained from Eq. \eqref{Jacobian}:
\begin{align}
\left[\mJ_{ij}^{\alpha\beta}\right]
&:=\left[ -\frac{\partial\tilde{F}_{i}^{\alpha}}{\partial q_{j}^{\beta}} \right]\nonumber \\
&=
\left[
\begin{matrix}
-\partial_{q_{j}^{x}}F_{i}^{x}    &  -\partial_{q_{j}^{y}}F_{i}^{x}  & -\partial_{q_{j}^{\ell}} F_{i}^{x} \\
-\partial_{q_{j}^{x}}F_{i}^{y}    &  -\partial_{q_{j}^{y}}F_{i}^{y}  & -\partial_{q_{j}^{\ell}} F_{i}^{y} \\
-\partial_{q_{j}^{x}}\tilde T_{i} & -\partial_{q_{j}^{y}}\tilde T_{i} & -\partial_{q_{j}^{\ell}}\tilde T_{i}
\end{matrix}
\right]
\nonumber \\
&=
\left[
\begin{matrix}
-\sum_{k=1;k\neq j}^{N}\partial_{q_{j}^{x}}f_{ik}^{x} & -\sum_{k=1;k\neq j}^{N}\partial_{q_{j}^{y}}f_{ik}^{x} & -\sum_{k=1;k\neq j}^{N}\partial_{q_{j}^{\ell}}f_{ik}^{x}\\
-\sum_{k=1;k\neq j}^{N}\partial_{q_{j}^{x}}f_{ik}^{y} & -\sum_{k=1;k\neq j}^{N}\partial_{q_{j}^{y}}f_{ik}^{y} & -\sum_{k=1;k\neq j}^{N}\partial_{q_{j}^{\ell}}f_{ik}^{y}\\
-\sum_{k=1;k\neq j}^{N}\partial_{q_{j}^{x}}\tilde T_{ik} & -\sum_{k=1;k\neq j}^{N}\partial_{q_{j}^{y}}\tilde T_{ik} & -\sum_{k=1;k\neq j}^{N}\partial_{q_{j}^{\ell}}\tilde T_{ik}
\end{matrix}
\right]
\nonumber \\
&=
\begin{cases}
\left[
\begin{matrix}
- \partial_{q_{j}^{x}}f_{ij}^{x} & - \partial_{q_{j}^{y}}f_{ij}^{x} & - \partial_{q_{j}^{\ell}}f_{ij}^{x}\\
- \partial_{q_{j}^{x}}f_{ij}^{y} & - \partial_{q_{j}^{y}}f_{ij}^{y} & - \partial_{q_{j}^{\ell}}f_{ij}^{y}\\
- \partial_{q_{j}^{x}}\tilde T_{ij} & - \partial_{q_{j}^{y}}\tilde T_{ij} & - \partial_{q_{j}^{\ell}}\tilde T_{ij}
\end{matrix}
\right]
(i\neq j)\\
\left[
\begin{matrix}
-\sum_{k=1;k\neq i}^{N}\partial_{q_{i}^{x}}f_{ik}^{x} & -\sum_{k=1;k\neq i}^{N}\partial_{q_{i}^{y}}f_{ik}^{x} & -\sum_{k=1;k\neq i}^{N}\partial_{q_{i}^{\ell}}f_{ik}^{x}\\
-\sum_{k=1;k\neq i}^{N}\partial_{q_{i}^{x}}f_{ik}^{y} & -\sum_{k=1;k\neq i}^{N}\partial_{q_{i}^{y}}f_{ik}^{y} & -\sum_{k=1;k\neq i}^{N}\partial_{q_{i}^{\ell}}f_{ik}^{y}\\
-\sum_{k=1;k\neq i}^{N}\partial_{q_{i}^{x}}\tilde T_{ik} & -\sum_{k=1;k\neq i}^{N}\partial_{q_{i}^{y}}\tilde T_{ik} & -\sum_{k=1;k\neq i}^{N}\partial_{q_{i}^{\ell}}\tilde T_{ik}
\end{matrix}
\right]
(i=j)
\end{cases}
\label{calJac}
,
\end{align}
where the superscripts $\alpha$ and $\beta$ correspond to $x,y,\ell$-components, and $i$ and $j$ are the particle numbers.
Here, $f_{ij}^{\zeta},\tilde T_{ij}$ are $\zeta$-component of $\bm{f}_{ij}$ and scaled torque that the $i$-th particle receives from the $j$-th particle, respectively.
The sub-matrix  for $i=j$ is given by
\begin{align}
\left[ \mJ_{ii}^{\alpha\beta} \right]
=
\left[
\begin{matrix}
\sum_{k=1;k\neq i}^{N}\partial_{q_{k}^{x}}f_{ik}^{x} & \sum_{k=1;k\neq i}^{N}\partial_{q_{k}^{y}}f_{ik}^{x} & -\sum_{k=1;k\neq i}^{N}\partial_{q_{k}^{\ell}}f_{ik}^{x}\\
\sum_{k=1;k\neq i}^{N}\partial_{q_{k}^{x}}f_{ik}^{y} & \sum_{k=1;k\neq i}^{N}\partial_{q_{k}^{y}}f_{ik}^{y} & -\sum_{k=1;k\neq i}^{N}\partial_{q_{k}^{\ell}}f_{ik}^{y}\\
\sum_{k=1;k\neq i}^{N}\partial_{q_{k}^{x}}\tilde T_{ik} & \sum_{k=1;k\neq i}^{N}\partial_{q_{k}^{y}}\tilde T_{ik} & -\sum_{k=1;k\neq i}^{N}\partial_{q_{k}^{\ell}}\tilde T_{ik}
\end{matrix}
\right]
\label{calJac2}
,
\end{align}
where we have used $\partial_{q_{i}^{\kappa}}f_{ik}^{\zeta} = -\partial_{q_{k}^{\kappa}}f_{ik}^{\zeta}, \partial_{q_{i}^{\kappa}}\tilde T_{ik} = -\partial_{q_{i}^{\kappa}}\tilde T_{ik}, \partial_{q_{i}^{\ell}}f_{ik}^{\zeta} = \partial_{q_{k}^{\ell}}f_{ik}^{\zeta}$, and $ \partial_{q_{i}^{\ell}}\tilde T_{ik} = \partial_{q_{i}^{\ell}}\tilde T_{ik}$.
Here, the superscripts $\zeta$ and $\kappa$ correspond to $x,y$ components.

\subsection{Derivation of Jacobian matrix in the harmonic system\label{AppHkn1}}
Let us consider only the normal and tangential elastic contact forces
\begin{align}
\bm{f}_{N,ij}&=k_{N}\xi_{N,ij}\bm{n}_{ij}, \label{Hertzian2}\\
\bm{f}_{T,ij}&=-k_{T}\bm{\xi}_{T,ij} \label{Hertzian3}
,
\end{align}
where the integration of $d\bm{\xi}_{T,ij}$
\begin{align}
\bm{\xi}_{T,ij}:=\int_{C_{ij}}d\bm{\xi}_{T,ij} \label{xiTint}
\end{align}
is performed during the contact between the $i$-th and $j$-th grains.
Since Eq. \eqref{xiTint} does not contain the second term on RHS of Eq. \eqref{xiT}, $\bm{\xi}_{T,ij}$ may not be perpendicular to $\bm{\xi}_{N,ij}$.
Nevertheless, we adopt Eq. \eqref{xiTint} for simplicity. 
Here, $d{\bm{\xi}}_{T,ij}$ is defined as
\begin{align}
d{\bm{\xi}}_{T,ij}= d{\bm{r}}_{ij}-(d{\bm{r}}_{ij}\cdot\bm{n}_{ij})\bm{n}_{ij}-d{\bm{\ell}}_{ij}\times\bm{n}_{ij}
\label{velotangential}
,
\end{align}
where $\bm{\ell}_{ij}$ is defined as
\begin{align}
\bm{\ell}_{ij}
:=
\left[
\begin{matrix}
0\\
0\\
\ell_{i}+\ell_{j}
\end{matrix}
\right].
\end{align}
Each component of Eq. \eqref{velotangential} is written as
\begin{align}
d{\xi}_{T,ij}^{x}
&= d{{r}}_{ij}^{x}-(d{\bm{r}}_{ij}\cdot\bm{n}_{ij}){n}_{ij}^{x}+d{{\ell}}_{ij}{n}_{ij}^{y}
, \\
d{\xi}_{T,ij}^{y}
&= d{{r}}_{ij}^{y}-(d{\bm{r}}_{ij}\cdot\bm{n}_{ij}){n}_{ij}^{y}-d{{\ell}}_{ij}{n}_{ij}^{x}
.
\end{align}

The derivative of the normal force is given by
\begin{align}
\partial_{r_{i}^{\zeta}}f_{N,ij}^{\kappa}
&=k_{N}\left[
\frac{\xi_{N,ij}}{r_{ij}}\delta_{\zeta\kappa} - \left( 1+\frac{\xi_{N,ij}}{r_{ij}} \right)n_{ij}^{\zeta}n_{ij}^{\kappa}
\right],
\label{N1st}
\\
\partial_{\ell_{i}}f_{N,ij}^{\kappa}&=0,
\end{align}
where Kronecker's delta $\delta_{\zeta\kappa}$ satisfies $\delta_{\zeta\kappa}=1$ for $\zeta=\kappa$ and $\delta_{\zeta\kappa}=0$ otherwise.
We have used
\begin{align}
\frac{\partial n_{ij}^{\zeta}}{\partial r_{i}^{\kappa}}&=\frac{1}{r_{ij}}\left( \delta_{\zeta\kappa}-n_{ij}^{\zeta}n_{ij}^{\kappa}\right),\\
\frac{\partial r_{ij}}{\partial r_{i}^{\zeta}}&=n_{ij}^{\zeta}
\end{align}
to obtain Eq. \eqref{N1st}.

The derivative of the tangential force is written as
\begin{align}
\partial_{r_{i}^{\zeta}}f_{T,ij}^{\kappa}
&= - k_{T}\left( \delta_{\zeta\kappa}-n_{ij}^{\zeta}n_{ij}^{\kappa} \right)
,
\label{T1st}
\\
\partial_{\ell_{i}}f_{T,ij}^{\kappa}
&=-\varepsilon_{\kappa}k_{T}n_{ij}^{\nu_{\kappa}}
\label{ellFt},
\end{align}
where $\varepsilon_{\zeta}$ and $\nu_{\zeta}$ are defined, respectively, as
\begin{align}
\varepsilon_{\zeta}:&=
\left\{
\begin{matrix}
1 \quad (\zeta=x)\\
-1 \quad (\zeta=y),
\end{matrix}
\right.
\\
\nu_{\zeta}:&=
\left\{
\begin{matrix}
y \quad (\zeta=x)\\
x \quad (\zeta=y).
\end{matrix}
\right.
\end{align}

Here, $\partial_{r_{i}^{\zeta}} {\xi}_{T,ij}^{\kappa}$ and $\partial_{\ell_{i}} {\xi}_{T,ij}^{\kappa} $ in Eqs. \eqref{T1st} and \eqref{ellFt} satisfy the following:
\begin{align}
\frac{\partial  {\xi}_{T,ij}^{\kappa}  }{\partial r_{i}^{\zeta}}
&=\delta_{\zeta\kappa}-n_{ij}^{\zeta}n_{ij}^{\kappa},\label{eq45}
\\
\frac{\partial  {\xi}_{T,ij}^{\kappa}  }{\partial \ell_{i}}
&=  \varepsilon_{\kappa} n_{ij}^{\nu_{\kappa}}
.
\label{eq46}
\end{align}
The derivation of Eqs. \eqref {eq45} and \eqref {eq46} are as follows \cite{Chattoraj19}.
From Eq. \eqref{velotangential}, $d {\xi}_{T,ij}^{\zeta}$ can be written as
\begin{align}
d {\xi}_{T,ij}^{\zeta}  = dr_{ij}^{\zeta}-(d\bm{r}_{ij}\cdot\bm{n}_{ij})n_{ij}^{\zeta} + (-1)^{\zeta}(d\ell_{i}+d\ell_{j})n_{ij}^{\nu_{\zeta}}
\label{eq23}
.
\end{align}
Then, $d\xi_{T,ij}^{x}$ satisfies
\begin{align}
d{\xi}_{T,ij}^{x} &= dr_{ij}^{x}-\sum_{\kappa=x,y}dr_{ij}^{\kappa}n_{ij}^{\kappa}n_{ij}^{x} + n_{ij}^{y}(d\ell_{i}+d\ell_{j})
\nonumber \\
&=(1-(n_{ij}^{x})^{2})dr_{ij}^{x} - n_{ij}^{x}n_{ij}^{y}dr_{ij}^{y} + n_{ij}^{y}(d\ell_{i}+d\ell_{j})
\nonumber \\
&= (n_{ij}^{y})^{2}dr_{ij}^{x} - n_{ij}^{x}n_{ij}^{y}dr_{ij}^{y} + n_{ij}^{y}(d\ell_{i}+d\ell_{j})
\nonumber \\
&= (n_{ij}^{y})^{2}(dx_{i}-dx_{j}) - n_{ij}^{x}n_{ij}^{y}(dy_{i}-dy_{j}) + n_{ij}^{y}(d\ell_{i}+d\ell_{j})
\label{dxitx}.
\end{align}
Similarly, $d\xi_{T,ij}^{y}$ also satisfies
\begin{align}
d\xi_{T,ij}^{y}
&= -n_{ij}^{x}n_{ij}^{y}(dx_{i}-dx_{j}) + (n_{ij}^{y})^{2}(dy_{i}-dy_{j}) - n_{ij}^{x}(d\ell_{i}+d\ell_{j})
\label{dxity}
.
\end{align}
Here, $d{\xi}_{T,ij}^{\zeta}$is the function of $x_{i}, y_{i}, \ell_{i}, x_j, y_j,$ and $\ell_{j}$.
We obtain the differential form of $d{\xi}_{T,ij}^{\zeta}$:
\begin{align}
d {\xi}_{T,ij}^{\zeta}
&=
  \left(\frac{\partial {\xi}_{T,ij}^{\zeta}}{\partial x_{i}}\right)_{(y_{i},\ell_{i},x_{j},y_{j},\ell_{j})}dx_{i}
+\left(\frac{\partial {\xi}_{T,ij}^{\zeta}}{\partial x_{j}}\right)_{(x_{i},y_{i},\ell_{i},y_{j},\ell_{j})}dx_{j}
\nonumber \\
&\quad
+\left(\frac{\partial {\xi}_{T,ij}^{\zeta}}{\partial y_{i}}\right)_{(x_{i},\ell_{i},x_{j},y_{j},\ell_{j})}dy_{i}
+\left(\frac{\partial {\xi}_{T,ij}^{\zeta}}{\partial y_{j}}\right)_{(x_{i},y_{i},\ell_{i},x_{j},\ell_{j})}dy_{j}
\nonumber \\
&\quad
+\left(\frac{\partial {\xi}_{T,ij}^{\zeta}}{\partial \ell_{i}}\right)_{(x_{i},y_{i},x_{j},y_{j},\ell_{j})}d\ell_{i}
+\left(\frac{\partial {\xi}_{T,ij}^{\zeta}}{\partial \ell_{j}}\right)_{(x_{i},y_{i},\ell_{i},x_{j},y_{j})}d\ell_{j}
\label{dxit}
.
\end{align}
Next, we obtain Eqs. \eqref{eq45}, \eqref{eq46}, by comparing Eqs. \eqref {dxitx} and \eqref {dxity} using Eq. \eqref {dxit}.

Because the scaled torque $\tilde T_{ij}$ satisfies
\begin{align}
\tilde{T}_{ij}:=\frac{2T_{ij}}{d_{i}}=-n_{ij}^{x}f_{T,ij}^{y} + n_{ij}^{y} f_{T,ij}^{x}
,
\end{align}
we obtain
\begin{align}
\partial_{r_{i}^{\zeta}}\tilde{T}_{ij}
&=-\left(\partial_{r_{i}^{\zeta}}n_{ij}^{x}\right)f_{T,ij}^{y} -n_{ij}^{x}\partial_{r_{i}^{\zeta}}f_{T,ij}^{y} + \left(\partial_{r_{i}^{\zeta}}n_{ij}^{y}\right) f_{T,ij}^{x} + n_{ij}^{y} \partial_{r_{i}^{\zeta}}f_{T,ij}^{x}
\nonumber \\
&=-\left( \frac{\delta_{\zeta x}}{r_{ij}}-\frac{n_{ij}^{\zeta}n_{ij}^{x}}{r_{ij}}\right)f_{T,ij}^{y}
+k_{T}n_{ij}^{x}
\left( \delta_{\zeta y}-n_{ij}^{\zeta}n_{ij}^{y} \right)
+\left( \frac{\delta_{\zeta y}}{r_{ij}}-\frac{n_{ij}^{\zeta}n_{ij}^{y}}{r_{ij}}\right)f_{T,ij}^{x}
-k_{T}n_{ij}^{y}\left( \delta_{\zeta x}-n_{ij}^{\zeta}n_{ij}^{x} \right)
\nonumber \\
&=-\varepsilon_{\kappa}\frac{n_{ij}^{\nu_{\kappa}}}{r_{ij}}\left(\bm{n}_{ij}\cdot\bm{f}_{T,ij}\right)
-\varepsilon_{\kappa}n_{ij}^{\nu_{\kappa}}\left( \bm{n}_{ij}\cdot\bm{n}_{ij} \right)
-\varepsilon_{\kappa}n_{ij}^{\nu_{\kappa}}
\label{xyT}
,
\\
\partial_{\ell_{i}}\tilde{T}_{ij}
&=-n_{ij}^{x}\partial_{\ell_{i}}f_{T,ij}^{y} + n_{ij}^{y}\partial_{\ell_{i}}f_{T,ij}^{x}
\nonumber \\
&=-n_{ij}^{x} k_{T}n_{ij}^{x}
-n_{ij}^{y} k_{T}n_{ij}^{y}
\nonumber \\
&=-k_{T}
\label{last}
,
\end{align}
where we have used $\bm{f}_{T,ij}\cdot \bm{n}_{ij}=0$ and $\bm{n}_{ij}\cdot\bm{n}_{ij}=1$.

\subsection{Explicit form of Jacobian for particles interacting with harmonic force \label{AppHkn2}}

From Sec.~\ref{AppHkn1}, the off-diagonal elements of the Jacobian matrix $\mJ_{ij}^{\alpha \beta}:=-\partial_{q_{j}^{s}}\tilde F_{i}^{r}\ (\alpha, \beta=x,y,\ell)$ with $i\ne j$ are given by
\begin{align}
\mJ_{ij}^{xx}
&=-k_{N}+k_{N}\left[ 1+\frac{\xi_{N,ij}}{r_{ij}}\right](n_{ij}^{y})^{2}
-k_{T}(n_{ij}^{y})^{2}
\label{Jxx}
,\\
\mJ_{ij}^{xy}
&=-k_{N}\left[ 1+\frac{\xi_{N,ij}}{r_{ij}}\right]n_{ij}^{x}n_{ij}^{y}
+k_{T}n_{ij}^{x}n_{ij}^{y}
,\\
\mJ_{ij}^{x\ell}
&=-k_{T}n_{ij}^{y}
,\\
\mJ_{ij}^{yx}
&=-k_{N}\left[ 1+\frac{\xi_{N,ij}}{r_{ij}}\right]n_{ij}^{x}n_{ij}^{y}
+k_{T}n_{ij}^{x}n_{ij}^{y}
,\\
\mJ_{ij}^{yy}
&=-k_{N}+k_{N}\left[ 1+\frac{\xi_{N,ij}}{r_{ij}}\right](n_{ij}^{x})^{2}
-k_{T}(n_{ij}^{x})^{2}
,\\
\mJ_{ij}^{y\ell}
&=k_{T}n_{ij}^{x}
,\\
\mJ_{ij}^{\ell x}
&=k_{T}n_{ij}^{y}
,\\
\mJ_{ij}^{\ell y}
&=-k_{T}n_{ij}^{x}
,\\
\mJ_{ij}^{\ell \ell}
&=k_{T} .
\label{Jll}
\end{align}
Notably, the elements of the Jacobian matrix are independent of $\xi_{T,ij}$.

With the aid of Eq.~\eqref{calJac}, the elements in the diagonal block $\mJ_{ii}^{rs}$ are given by
\begin{align}
\mJ_{ii}^{xx}
&=-\sum_{j\neq i}\left\{
-k_{N}+k_{N}\left[ 1+\frac{\xi_{N,ij}}{r_{ij}}\right](n_{ij}^{y})^{2}
-k_{T}(n_{ij}^{y})^{2}
\right\}
,\\
\mJ_{ii}^{xy}
&=-\sum_{j\neq i}\left\{
-k_{N}\left[ 1+\frac{\xi_{N,ij}}{r_{ij}}\right]n_{ij}^{x}n_{ij}^{y}
+k_{T}n_{ij}^{x}n_{ij}^{y}
\right\}
,\\
\mJ_{ii}^{x\ell}
&=\sum_{j\neq i}k_{T}n_{ij}^{y}
,\\
\mJ_{ii}^{yx}
&=-\sum_{j\neq i}\left\{
-k_{N}\left[ 1+\frac{\xi_{N,ij}}{r_{ij}}\right]n_{ij}^{x}n_{ij}^{y}
+k_{T}n_{ij}^{x}n_{ij}^{y}
\right\}
,\\
\mJ_{ii}^{yy}
&=-\sum_{j\neq i}\left\{
-k_{N}+k_{N}\left[ 1+\frac{\xi_{N,ij}}{r_{ij}}\right](n_{ij}^{x})^{2}
-k_{T}(n_{ij}^{x})^{2}
\right\}
,\\
\mJ_{ii}^{y\ell}
&=-\sum_{j\neq i}k_{T}n_{ij}^{x}
,\\
\mJ_{ii}^{\ell x}
&=\sum_{j\neq i}k_{T}n_{ij}^{y}
,\\
\mJ_{ii}^{\ell y}
&=-\sum_{j\neq i}k_{T}n_{ij}^{x}
,\\
\mJ_{ii}^{\ell\ell}
&=\sum_{j\neq i}k_{T} .
\label{J_ii^ll}
\end{align}
The expressions in Eqs.~\eqref{Jxx}-\eqref{J_ii^ll} are equivalent to Eqs.~\eqref{xxH}-\eqref{llH} for a Hessian matrix. 
Thus, we conclude that the Jacobian matrix is equivalent to the Hessian matrix for the harmonic system without considering Coulomb's slip.

\section{The detailed properties of rigidity}\label{app:rigidity}

This appendix consists of two sections.
In Appendix \ref{xxx}, we present the detailed expressions of rigidity $g$.
In Appendix \ref{AppHkn4}, we demonstrate the quantitative accuracy of the Hessian analysis in the linear response regime by comparing  the theoretical evaluation of the rigidity with that obtained by the numerical simulation as in Ref.~\cite{Ishima2022}.

\subsection{The expression of $g$ \label{xxx}}
The symmetrized shear stress in Eq.~\eqref{Sigma} is expressed as
\begin{align}
\sigma(\bm{q}^{\textrm{FB}}(\gamma))=\frac{\sigma_{xy}(\bm{q}^{\textrm{FB}}(\gamma))+\sigma_{yx}(\bm{q}^{\textrm{FB}}(\gamma))}{2} ,
\label{SgmUseTnsr}
\end{align}
where
\begin{align}
\sigma_{\alpha\beta}(\bm{q}^{\textrm{FB}}(\gamma))
:=-\frac{1}{2L^{2}} \sum_{i,j(i\neq j)}
f_{ij}^{\alpha}(\bm{q}^{\textrm{FB}}(\gamma)) r_{ij}^{\beta}(\bm{q}^{\textrm{FB}}(\gamma)) .
\label{StressTensor}
\end{align}
Substituting Eq.~\eqref{SgmUseTnsr} into Eq.~\eqref{Gnm}, we obtain
\begin{align}
g(\gamma)
&=\lim_{\Delta\gamma\to0}
\frac{1}{2\Delta\gamma}
\left[
\sigma_{xy}(\bm{q}^{\textrm{FB}}(\gamma+\Delta\gamma))+\sigma_{yx}(\bm{q}^{\textrm{FB}}(\gamma+\Delta\gamma))
-
\left\{
\sigma_{xy}(\bm{q}^{\textrm{FB}}(\gamma))+\sigma_{yx}(\bm{q}^{\textrm{FB}}(\gamma))
\right\}
\right]
\nonumber \\
&=-\lim_{\Delta\gamma\to0}
\frac{1}{4\Delta\gamma L^{2}}
\left[
f_{ij}^{x}(\bm{q}^{\textrm{FB}}(\gamma+\Delta\gamma)) r_{ij}^{y}(\bm{q}^{\textrm{FB}}(\gamma+\Delta\gamma))
+f_{ij}^{y}(\bm{q}^{\textrm{FB}}(\gamma+\Delta\gamma)) r_{ij}^{x}(\bm{q}^{\textrm{FB}}(\gamma+\Delta\gamma))
\right.
\nonumber \\
&\quad\quad\quad\quad\quad\quad\quad\quad\quad
\left.
-
\left\{
f_{ij}^{x}(\bm{q}^{\textrm{FB}}(\gamma)) r_{ij}^{y}(\bm{q}^{\textrm{FB}}(\gamma))
+f_{ij}^{y}(\bm{q}^{\textrm{FB}}(\gamma)) r_{ij}^{x}(\bm{q}^{\textrm{FB}}(\gamma))
\right\}
\right] .
\label{gApp}
\end{align}
Substituting Eq.~\eqref{gApp} into Eq.~\eqref{SgmUseTnsr}, the rigidity $g$ is expressed as
\begin{align}
g(\gamma)&=-
\lim_{\Delta\gamma\to0}
\frac{1}{4\Delta\gamma L^{2}}
\left[
f_{ij}^{x}(\bm{q}^{\textrm{FB}}(\gamma+\Delta\gamma)) r_{ij}^{y}(\bm{q}^{\textrm{FB}}(\gamma+\Delta\gamma))
+f_{ij}^{y}(\bm{q}^{\textrm{FB}}(\gamma+\Delta\gamma)) r_{ij}^{x}(\bm{q}^{\textrm{FB}}(\gamma+\Delta\gamma))
\right.
\nonumber \\
&\quad\quad\quad\quad\quad\quad\quad\quad\quad
\left.
-
\left\{
f_{ij}^{x}(\bm{q}^{\textrm{FB}}(\gamma)) r_{ij}^{y}(\bm{q}^{\textrm{FB}}(\gamma))
+f_{ij}^{y}(\bm{q}^{\textrm{FB}}(\gamma)) r_{ij}^{x}(\bm{q}^{\textrm{FB}}(\gamma))
\right\}
\right].
\label{GdefApp}
\end{align}

By expanding $r_{ij}^{\alpha}(\bm{q}^{\textrm{FB}}(\gamma+\Delta\gamma))$ in Eq. \eqref{GdefApp} by $\Delta\gamma$ from the finite strain $\gamma$, 
we obtain
\begin{align}
r_{ij}^{\alpha}(\bm{q}^{\textrm{FB}}(\gamma+\Delta\gamma))
&=r_{i}^{\alpha}(\bm{q}^{\textrm{FB}}(\gamma+\Delta\gamma))-r_{j}^{\alpha}(\bm{q}^{\textrm{FB}}(\gamma+\Delta\gamma)) \nonumber \\
&\simeq r_{ij}^{\alpha}(\bm{q}^{\textrm{FB}}(\gamma)) + \Delta\gamma\left\{ \delta_{\alpha x}\left(y_{i}(\bm{q}^{\textrm{FB}}(\gamma))-y_{j}(\bm{q}^{\textrm{FB}}(\gamma))\right) + \frac{d\mathring r_{i}^{\alpha}(\bm{q}^{\textrm{FB}}(\gamma))}{d\gamma} - \frac{d\mathring r_{j}^{\alpha}(\bm{q}^{\textrm{FB}}(\gamma))}{d\gamma} \right\} \nonumber \\
&=r_{ij}^{\alpha}(\bm{q}^{\textrm{FB}}(\gamma)) + \Delta\gamma\left\{ \delta_{\alpha x}y_{ij}(\bm{q}^{\textrm{FB}}(\gamma))+\frac{d\mathring r_{ij}^{\alpha}(\bm{q}^{\textrm{FB}}(\gamma))}{d\gamma} \right\}
\label{rij}
.
\end{align}

Similarly, by expanding $f_{ij}^{\alpha}(\gamma+\Delta\gamma)$ in Eq. \eqref{GdefApp} from the zero strain state, we obtain
\begin{align}
f_{ij}^{\alpha}(\bm{q}^{\textrm{FB}}(\gamma+\Delta\gamma))
&\simeq f_{ij}^{\alpha}(\bm{q}^{\textrm{FB}}(\gamma))
+ \sum_{k=1}^{N}\sum_{\zeta=x,y}\Delta\gamma\frac{\partial f_{ij}^{\alpha}}{\partial r_{k}^{\zeta}}\frac{d r_{k}^{\zeta}}{d\gamma}
+ \sum_{k=1}^{N}\Delta\gamma\frac{\partial f_{ij}^{\alpha}}{\partial \ell_{k}}\frac{d \ell_{k}}{d\gamma}
 \nonumber \\
&=f_{ij}^{\alpha}(\bm{q}^{\textrm{FB}}(\gamma))
+ \sum_{\zeta=x,y}\Delta\gamma\left[
\frac{\partial f_{ij}^{\alpha}}{\partial r_{i}^{\zeta}}\Biggl( \delta_{\zeta x}y_{i}(\bm{q}^{\textrm{FB}}(\gamma))+\frac{d\mathring r_{i}^{\zeta}(\bm{q}^{\textrm{FB}}(\gamma))}{d\gamma} \right)
+
\frac{\partial f_{ij}^{\alpha}}{\partial r_{j}^{\zeta}}\left( \delta_{\zeta x}y_{j}(\bm{q}^{\textrm{FB}}(\gamma))+\frac{d\mathring r_{j}^{\zeta}(\bm{q}^{\textrm{FB}}(\gamma))}{d\gamma} \Biggl)
\right]
\nonumber \\
&\quad
+ \Delta\gamma\left[
\frac{\partial f_{ij}^{\alpha}}{\partial \ell_{i}}\Biggl( \delta_{\ell x}y_{i}(\bm{q}^{\textrm{FB}}(\gamma))+\frac{d\mathring \ell_{i}(\bm{q}^{\textrm{FB}}(\gamma))}{d\gamma} \right)
+
\frac{\partial f_{ij}^{\alpha}}{\partial \ell_{j}}\left( \delta_{\ell x}y_{j}(\bm{q}^{\textrm{FB}}(\gamma))+\frac{d\mathring \ell_{j}(\bm{q}^{\textrm{FB}}(\gamma))}{d\gamma} \Biggl)
\right]
.
\end{align}
Moreover, using ${\partial f_{ij}^{\alpha}}/{\partial r_{j}^{\zeta}}=-{\partial f_{ij}^{\alpha}}/{\partial r_{i}^{\zeta}}$ and $ {\partial f_{ij}^{\alpha}}/{\partial \ell_{j}}={\partial f_{ij}^{\alpha}}/{\partial \ell_{i}}$, and $f_{ij}^{\alpha}$ can be written as
\begin{align}
f_{ij}^{\alpha}(\bm{q}^{\textrm{FB}}(\gamma+\Delta\gamma))
&=f_{ij}^{\alpha}(\bm{q}^{\textrm{FB}}(\gamma))
+\sum_{\zeta=x,y} \Delta\gamma \frac{\partial f_{ij}^{\alpha}}{\partial r_{i}^{\zeta}}\left( \delta_{\zeta x}y_{ij}(\bm{q}^{\textrm{FB}}(\gamma))+\frac{d\mathring r_{ij}^{\zeta}(\bm{q}^{\textrm{FB}}(\gamma))}{d\gamma} \right)
+ \Delta\gamma \frac{\partial f_{ij}^{\alpha}}{\partial \ell_{i}}  \left(\frac{d\mathring \ell_{i}(\bm{q}^{\textrm{FB}}(\gamma))}{d\gamma} + \frac{d\mathring \ell_{j}(\bm{q}^{\textrm{FB}}(\gamma))}{d\gamma} \right)
\label{fij}
.
\end{align}

Substituting Eqs. \eqref{rij} and \eqref{fij} into Eq. \eqref{GdefApp}, we obtain
\begin{align}
g(\gamma)
&=
-\frac{1}{4 L^{2}}
\sum_{i,j(i\neq j)}
\Biggl[
f_{ij}^{x}(\bm{q}^{\textrm{FB}}(\gamma))\frac{d\mathring r_{ij}^{y}(\bm{q}^{\textrm{FB}}(\gamma))}{d \gamma}
+
f_{ij}^{y}(\bm{q}^{\textrm{FB}}(\gamma))\frac{d\mathring r_{ij}^{x}(\bm{q}^{\textrm{FB}}(\gamma))}{d \gamma}
\nonumber \\
&\quad\quad\quad\quad\quad
+\sum_{\zeta=x,y}
\left(
\frac{\partial f_{ij}^{x}(\bm{q}^{\textrm{FB}}(\gamma))}{\partial r_{i}^{\zeta}}r_{ij}^{y}(\bm{q}^{\textrm{FB}}(\gamma))
+
\frac{\partial f_{ij}^{y}(\bm{q}^{\textrm{FB}}(\gamma))}{\partial r_{i}^{\zeta}}r_{ij}^{x}(\bm{q}^{\textrm{FB}}(\gamma)) 
\right)
\left( \delta_{\zeta x}y_{ij}(\bm{q}^{\textrm{FB}}(\gamma))+\frac{d\mathring r_{ij}^{\zeta}(\bm{q}^{\textrm{FB}}(\gamma))}{d\gamma} \right)
\nonumber \\
&\quad\quad\quad\quad\quad
+
\left(
\frac{\partial f_{ij}^{x}(\bm{q}^{\textrm{FB}}(\gamma))}{\partial \ell_{i}}r_{ij}^{y}(\bm{q}^{\textrm{FB}}(\gamma))
+
\frac{\partial f_{ij}^{y}(\bm{q}^{\textrm{FB}}(\gamma))}{\partial \ell_{i}}r_{ij}^{xx}(\bm{q}^{\textrm{FB}}(\gamma))
\right)
\left(\frac{d\mathring \ell_{i}(\bm{q}^{\textrm{FB}}(\gamma))}{d\gamma} + \frac{d\mathring \ell_{j}(\bm{q}^{\textrm{FB}}(\gamma))}{d\gamma}\right)
\Biggl]
\label{eq52}
.
\end{align}
Because $\sum_{i(i\neq j)}f_{ij}^{\alpha}(\bm{q}^{\textrm{FB}}(\gamma))=0$ in the FB state, 
the first and second terms on the RHS of Eq. \eqref{eq52} can be written as
\begin{align}
\sum_{i,j(i\neq j)}f_{ij}^{\alpha}(\bm{q}^{\textrm{FB}}(\gamma))\frac{d\mathring{r}_{ij}^{\kappa}(\bm{q}^{\textrm{FB}}(\gamma))}{d\gamma}
&=\sum_{i,j(i\neq j)}f_{ij}^{\alpha}(\bm{q}^{\textrm{FB}}(\gamma))\left( \frac{d\mathring{r}_{i}^{\kappa}(\bm{q}^{\textrm{FB}}(\gamma))}{d\gamma} - \frac{d\mathring{r}_{j}^{\kappa}(\bm{q}^{\textrm{FB}}(\gamma))}{d\gamma}\right)
\nonumber \\
&=\sum_{j}\left(\sum_{j(j\neq i)}f_{ij}^{\alpha}(\bm{q}^{\textrm{FB}}(\gamma))\right) \frac{d\mathring{r}_{i}^{\kappa}(\bm{q}^{\textrm{FB}}(\gamma))}{d\gamma}
-\sum_{i}\left(\sum_{i(i\neq j)} f_{ij}^{\alpha}(\bm{q}^{\textrm{FB}}(\gamma))\right) \frac{d\mathring{r}_{j}^{\kappa}(\bm{q}^{\textrm{FB}}(\gamma))}{d\gamma}
\nonumber \\
&=0
\label{eq53}
.
\end{align}
Thus, $g$ is expressed as
 \begin{align}
 g(\gamma)
&=
-\frac{1}{4 L^{2}}
\sum_{i,j(i\neq j)}
\Biggl[
\sum_{\zeta=x,y}
\left(
\frac{\partial f_{ij}^{x}(\bm{q}^{\textrm{FB}}(\gamma))}{\partial r_{i}^{\zeta}}r_{ij}^{y}(\bm{q}^{\textrm{FB}}(\gamma))
+
\frac{\partial f_{ij}^{y}(\bm{q}^{\textrm{FB}}(\gamma))}{\partial r_{i}^{\zeta}}r_{ij}^{x}(\bm{q}^{\textrm{FB}}(\gamma)) 
\right)
\left( \delta_{\zeta x}y_{ij}(\bm{q}^{\textrm{FB}}(\gamma))+\frac{d\mathring r_{ij}^{\zeta}(\bm{q}^{\textrm{FB}}(\gamma))}{d\gamma} \right)
\nonumber \\
&\quad\quad\quad\quad\quad
+
\left(
\frac{\partial f_{ij}^{x}(\bm{q}^{\textrm{FB}}(\gamma))}{\partial \ell_{i}}r_{ij}^{y}(\bm{q}^{\textrm{FB}}(\gamma))
+
\frac{\partial f_{ij}^{y}(\bm{q}^{\textrm{FB}}(\gamma))}{\partial \ell_{i}}r_{ij}^{xx}(\bm{q}^{\textrm{FB}}(\gamma))
\right)
\left(\frac{d\mathring \ell_{i}(\bm{q}^{\textrm{FB}}(\gamma))}{d\gamma} + \frac{d\mathring \ell_{j}(\bm{q}^{\textrm{FB}}(\gamma))}{d\gamma}\right)
\Biggl]
.
\end{align}
Using $\mH_{ij}^{\alpha\beta}=\mJ_{ij}^{\alpha\beta}:=-\partial_{q_{j}^{\beta}}f_{ij}^{\alpha}$
for $i\neq j$ in the case of the harmonic contact model, 
we can express $g$ as
\begin{align}
g(\gamma)
&=
\frac{1}{4L^{2}}
\sum_{i,j(i\neq j)}
\left[
\sum_{\zeta=x,y}
\left( y_{ij}(\bm{q}^{\textrm{FB}}(\gamma))\mH_{ji}^{x\zeta}(\bm{q}^{\textrm{FB}}(\gamma)) + x_{ij}(\bm{q}^{\textrm{FB}}(\gamma))\mH_{ji}^{y\zeta}(\bm{q}^{\textrm{FB}}(\gamma)) \right)
\left( \delta_{\zeta x}y_{ij}(\bm{q}^{\textrm{FB}}(\gamma))+\frac{d\mathring r_{ij}^{\zeta}(\bm{q}^{\textrm{FB}}(\gamma))}{d\gamma} \right)
\right.
\nonumber \\
&\quad\quad\quad\quad\quad\quad\quad
\left.
+\left(y_{ij}(\bm{q}^{\textrm{FB}}(\gamma))\mH_{ji}^{x\ell}(\bm{q}^{\textrm{FB}}(\gamma)) + x_{ij}(\bm{q}^{\textrm{FB}}(\gamma))\mH_{ji}^{y\ell}(\bm{q}^{\textrm{FB}}(\gamma)) \right)
\left( \frac{d\mathring \ell_{i}(\bm{q}^{\textrm{FB}}(\gamma))}{d\gamma} + \frac{d\mathring \ell_{j}(\bm{q}^{\textrm{FB}}(\gamma))}{d\gamma} \right)
\right]
\nonumber \\
&=
\frac{1}{4L^{2}}
\sum_{i,j(i\neq j)}
\Biggl[
y_{ij}\left( y_{ij}(\bm{q}^{\textrm{FB}}(\gamma))\mH_{ji}^{xx}(\bm{q}^{\textrm{FB}}(\gamma)) + x_{ij}(\bm{q}^{\textrm{FB}}(\gamma))\mH_{ji}^{yx}(\bm{q}^{\textrm{FB}}(\gamma)) \right)
\nonumber \\
&\quad\quad\quad\quad
+\sum_{\zeta=x,y}\left( y_{ij}(\bm{q}^{\textrm{FB}}(\gamma))\mH_{ji}^{x\zeta}(\bm{q}^{\textrm{FB}}(\gamma)) + x_{ij}(\bm{q}^{\textrm{FB}}(\gamma))\mH_{ji}^{y\zeta}(\bm{q}^{\textrm{FB}}(\gamma)) \right) \frac{d\mathring r_{ij}^{\zeta}(\bm{q}^{\textrm{FB}}(\gamma))}{d\gamma}
\nonumber \\
&\quad\quad\quad\quad\quad\quad\quad
+\left( y_{ij}(\bm{q}^{\textrm{FB}}(\gamma))\mH_{ji}^{x\ell}(\bm{q}^{\textrm{FB}}(\gamma)) + x_{ij}(\bm{q}^{\textrm{FB}}(\gamma))\mH_{ji}^{y\ell}(\bm{q}^{\textrm{FB}}(\gamma))
\right) \left( \frac{d\mathring \ell_{i}(\bm{q}^{\textrm{FB}}(\gamma))}{d\gamma} + \frac{d\mathring \ell_{j}(\bm{q}^{\textrm{FB}}(\gamma))}{d\gamma} \right)
\Biggl]
\label{eqG}
.
\end{align}
Thus, using Eqs \eqref{tildeJ1} and \eqref{tildeJ2},
we obtain Eqs. \eqref{expG}-\eqref{GNAff}.

\subsection{The rigidity of the harmonic model in the linear response regime}\label{AppHkn4}

In this appendix, we verify the validity of our method to evaluate the rigidity $G_{0}$ for frictional harmonic grains in the linear response regime for various $k_{T}/k_{N}$ and $\phi$ as in Ref.~\cite{Ishima2022}.
Figure~\ref{G0Lin} presents the results of the rigidity, in which $G_0$ obtained by the eigenvalue analysis (filled symbols) is in perfect agreement with that obtained by the simulation (open symbols).
Here we take the average over 5 ensembles for each $\phi$ and $k_T$. 
This figure confirms the validity of the theoretical method in the linear response regime.

\begin{figure}[htbp]
\centering
\includegraphics[width=8cm]{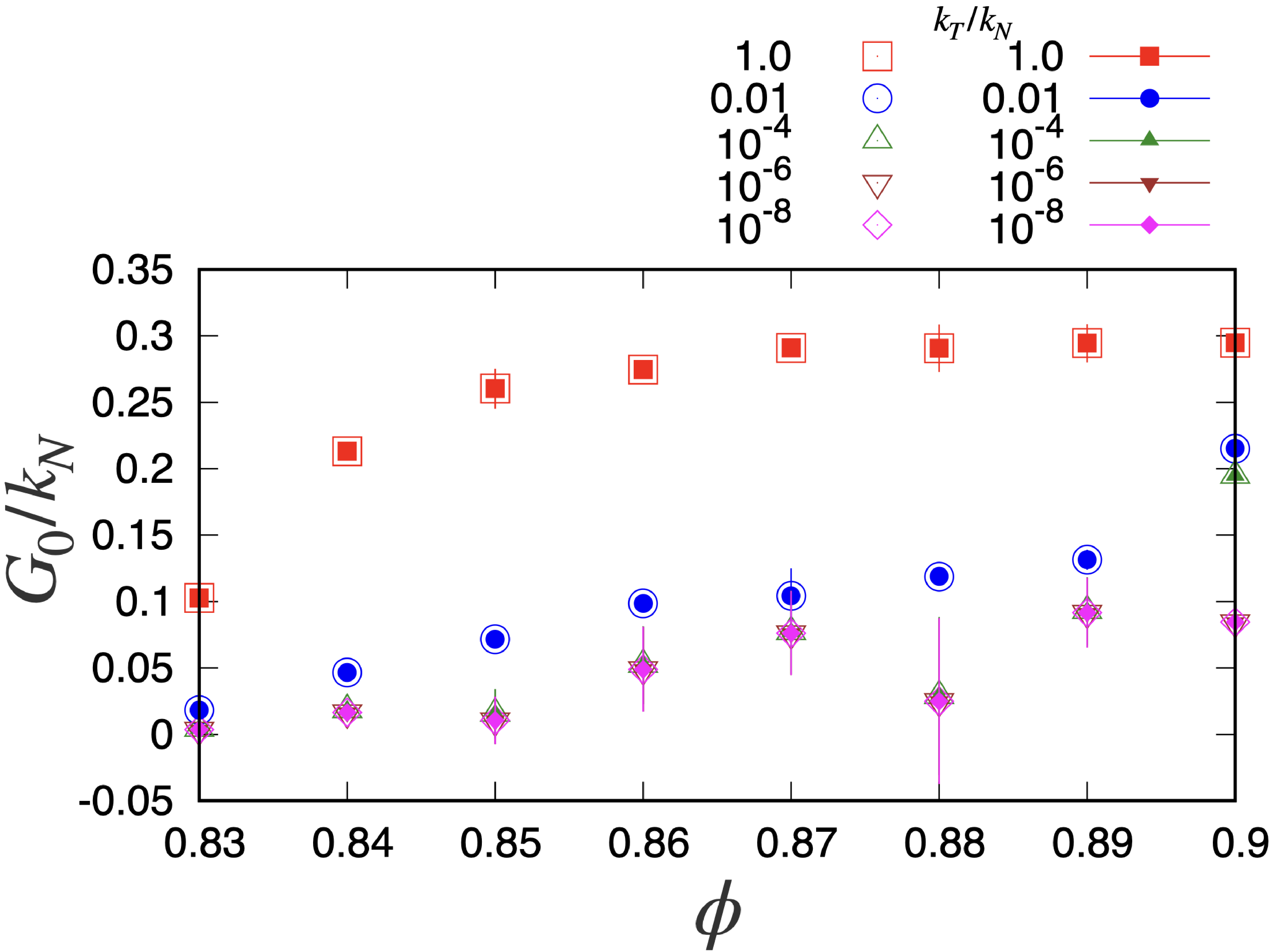}
\caption{
The plot of the rigidity $G_{0}$ in the linear response regime for various $k_{T}/k_{N}$ and $\phi$, 
where open symbols and filled symbols are results of the theory and simulation, respectively.
}
\label{G0Lin}
\end{figure}

\end{widetext}


\end{document}